\documentclass[acmsmall]{acmart}

\usepackage{xspace}
\usepackage{graphicx}
\usepackage{subfig}
\usepackage{wrapfig}
\usepackage{enumitem}
\usepackage{listings}

\AtBeginDocument{%
  }

\setcopyright{cc}
\setcctype{by-sa}
\acmJournal{PACMMOD}
\acmYear{2025} \acmVolume{3} \acmNumber{3 (SIGMOD)} \acmArticle{200} \acmMonth{6}
\acmPrice{}\acmDOI{10.1145/3725337}

\newcommand{\sys}{\textsc{Pneuma}\xspace}
\newcommand{\fts}{\textsc{Full-Text Search}\xspace}
\newcommand{\rag}{\textsc{LlamaIndex}\xspace}
\newcommand{\solo}{\textsc{Solo}\xspace}

\newcommand{\tinyskip}{\addvspace{3pt}}
\newcommand{\mypar}[1]{\tinyskip\noindent\textbf{#1.}\xspace}
\newcommand{\myparnoperiod}[1]{\tinyskip\noindent\textbf{#1}}

\newenvironment{myitemize}{
\begin{itemize}[leftmargin=1em, itemsep=.1em, parsep=.1em, topsep=.1em,
    partopsep=.1em]}
{\end{itemize}}
\newenvironment{myenumerate}{
\begin{enumerate}[leftmargin=1em, itemsep=.1em, parsep=.1em, topsep=.1em,
    partopsep=.1em]}
{\end{enumerate}}

\newenvironment{structure*}{\color{blue}\begin{myenumerate}}{\end{myenumerate}}

\lstdefinestyle{sqlStyle}{
    language=SQL,
    basicstyle=\footnotesize\ttfamily,
    keywordstyle=\color{blue}\bfseries,
    stringstyle=\color{red},
    commentstyle=\color{green},
    breaklines=true
}

\newtheorem{defn}{Definition}[section]

\newtheorem{problem}[defn]{Problem}

\begin{document}
\title{\sys: Leveraging LLMs for Tabular Data Representation and Retrieval in an End-to-End System}
\newcommand{\uchicago}{
  \institution{The University of Chicago}
  \city{Chicago}
  \country{USA}
}
\newcommand{\ui}{
  \institution{University of Indonesia}
  \city{Depok}
  \country{Indonesia}
}
\author{Muhammad Imam Luthfi Balaka}
\email{muhammad.imam07@ui.ac.id}
\orcid{0009-0001-5324-7758}
\affiliation{\ui}

\author{David Alexander}
\email{david.alexander01@ui.ac.id}
\orcid{0009-0008-6239-9092}
\affiliation{\ui}

\author{Qiming Wang}
\email{qmwang@uchicago.edu}
\orcid{0009-0004-6554-4161}
\affiliation{\uchicago}

\author{Yue Gong}
\email{yuegong@uchicago.edu}
\orcid{0009-0002-8646-473X}
\affiliation{\uchicago}

\author{Adila Krisnadhi}
\email{adila@cs.ui.ac.id}
\orcid{0000-0003-0745-6804}
\affiliation{\ui}

\author{Raul Castro Fernandez}
\email{raulcf@uchicago.edu}
\orcid{0000-0001-7675-6080}
\affiliation{\uchicago}

\renewcommand{\shortauthors}{Balaka et al.}

\begin{abstract}

Finding relevant tables among databases, lakes, and repositories is the first step in extracting value from data. Such a task remains difficult because assessing whether a table is relevant to a problem does not always depend only on its content but also on the context, which is usually tribal knowledge known to the individual or team. While tools like data catalogs and academic data discovery systems target this problem, they rely on keyword search or more complex interfaces, limiting non-technical users’ ability to find relevant data. The advent of large language models (LLMs) offers a unique opportunity for users to ask questions directly in natural language, making dataset discovery more intuitive, accessible, and efficient.

In this paper, we introduce \sys{}, a retrieval-augmented generation (RAG) system designed to efficiently and effectively discover tabular data. \sys{} leverages large language models (LLMs) for both table representation and table retrieval. For table representation, \sys{} preserves schema and row-level information to ensure comprehensive data understanding. For table retrieval, \sys{} augments LLMs with traditional information retrieval techniques, such as full-text and vector search, harnessing the strengths of both to improve retrieval performance. To evaluate \sys{}, we generate comprehensive benchmarks that simulate table discovery workload on six real-world datasets including enterprise data, scientific databases, warehousing data, and open data. Our results demonstrate that \sys{} outperforms widely used table search systems (such as full-text search and state-of-the-art RAG systems) in accuracy and resource efficiency. 

\end{abstract}



% Keywords

\received{October 2024}
\received[revised]{January 2025}
\received[accepted]{February 2025}

\maketitle

\section{Introduction}
\label{sec:introduction}

Identifying \emph{relevant} data is a prerequisite to solving data problems and creating value. Data may be relevant to a data problem because of its \emph{content}, i.e., columns and rows in tabular data. For instance, to calculate total sales for a product, a relevant table must include columns related to sales and products, with rows containing individual sales records. Data may also be relevant to a data problem because of its \emph{context}~\cite{gebru2021datasheets}, e.g., an enterprise data catalog that contains the origin, purpose and usage of the datasets. For example, to fill in missing values, it is crucial to understand the mechanism that explains the missing data~\cite{emmanuel2021survey}. Since context is often not explicitly included in table content, both content and context are commonly considered when searching for data~\cite{chen2020open}.

% ; this information is rarely included in the table content. Furthermore, it is common to refer to both the content and context when searching for data~\cite{chen2020open}.

Consider a user who asks for temperature data from a room that was sampled uniformly, e.g., because they need to know the sampling method for some downstream analysis. It is possible that a schema explicitly indicates the sampling procedure for the data. Still, it is also plausible that this information is instead incorporated into some form of documentation, e.g., a PDF, associated with the table. As an example, of 300k datasets in data.gov, $\approx$150k contain some form of associated context (in PDF, Text, or HTML). Our experience with organizations reveals that internal data is similarly documented in external repositories such as wikis, catalogs, and others~\cite{metis, amundsen}. Ideally, the user articulates their need without thinking whether satisfying that need requires looking at the content of the table---a schema that models sampling method---or context, the PDF. Today's solutions consider content and context when computing relevance. Data discovery systems largely overlook context~\cite{fernandez2018aurum, herzig2021open, wang2021retrieving, wang2023solo} while data catalogs index context. Furthermore, keyword search remains the primary interface in many dataset search platforms, with few offering natural language interfaces to address these types of data discovery tasks~\cite{wang2023solo}, yet natural language remains the most accessible method for non-technical users to identify relevant data.

In this paper, we present \sys{},\footnote{code is available at \url{https://github.com/TheDataStation/pneuma}} an open-source system that retrieves tables from table repositories based on their content, context, or both. After indexing a collection of tables, \sys{} takes natural language questions as input and produces a ranking of relevant tables as output. A main challenge \sys{} faces is that content and context are represented fundamentally differently. While the content of tables is highly structured, the context representation may range from free text to different degrees of semi-structuredness. To address this, \sys{} leverages large language models (LLMs) within a retrieval-augmented generation (RAG) architecture~\cite{lewis2020retrieval}, enabling both content and context to be represented as vectors. This paper contributes to the fields of data discovery and data catalogs by exploring two primary questions that arise from \sys{}'s approach:

\myparnoperiod{1. How do we represent content and context as vectors?} There are numerous tabular representation techniques~\cite{zhang2018ad, herzig2021open, wang2021retrieving, yin2020tabert, wang2023solo}, with many focused on generating vectors to achieve high retrieval accuracy. While accuracy is crucial, we find that these methods often produce indices with storage footprints several orders of magnitude larger than the original data, severely impacting scalability and making them impractical for large data collections.

Instead, \sys{} introduces a novel table representation method that leverages LLMs to \emph{narrate} table schema. LLMs, equipped with extensive knowledge, can provide meaningful column descriptions, even for abbreviations or domain-specific terms that may be challenging for humans or smaller models to interpret. For example, one table in our benchmarks has the schema \textsf{<Player | AB | AVG | BABIP>}, which might be ambiguous to a user unfamiliar with baseball statistics, but an LLM recognizes these as Major League Baseball (MLB) statistics. AB refers to ``the number of at-bats for a batter'' and BABIP to ``Batting Average on Balls In Play''. Schema narrations are further complemented by selected portions of the table's content, helping resolve semantic ambiguities and providing a more comprehensive table representation. \sys{}'s table representation method handles both content and context by transforming them into free text as an intermediate step. This text is then encoded into vectors using state-of-the-art embedding techniques~\cite{MTEB2023}.

\myparnoperiod{2. How do we ensure high retrieval accuracy?} To return relevant tables, the retrieval method must be compatible with the underlying vector representation. Retrieval techniques like full-text search~\cite{robertson2009probabilistic} excel when string matching\footnote{or simple approximate string matching} suffices but suffer whenever there is semantic ambiguity, like when the input natural question uses a language different than the tables. Vector search handles semantic ambiguity better~\cite{karpukhin2020dense}, but alone, it rarely offers high-quality answers if test data is beyond the domain of training data. RAG architectures treat vector search as a process to generate candidates which are then processed by a downstream LLM~\cite{gao2023retrieval}, i.e., the vector search results are incorporated into the LLM's context window. 

\sys{} introduces a retrieval method that improves over the above. It combines full-text and vector search, \emph{and} leverages an LLM as a mechanism to \emph{refine} the suggested ranking, rather than relying on the LLM to directly rank or answer the question. In other words, \sys{}'s retriever takes on more responsibility during candidate generation, reducing the workload for the downstream LLM. The LLM’s role is simplified to identifying and filtering out irrelevant entries from the final ranking. Since the LLM is tasked with a more focused, manageable job---but one that would still be difficult for traditional methods to handle---it performs much better than with more open-ended tasks. We find that this combination of retrieval and LLM evaluation significantly improves retrieval accuracy for both content and context questions.

In addition to these two primary technical contributions, this paper contributes two artifacts, a benchmark generator and an end-to-end system that implements the above techniques. Although data contexts must be supplied by external agents in practice, determining how to incentivize individuals to provide these contexts is beyond the scope of this paper. Instead, we simulate data contexts in our benchmark generator using an LLM. The benchmark generator takes table collections as input and produces content and context questions. It leverages LLMs to generate meaningful content questions and to create table contexts by answering the questions introduced in \cite{gebru2021datasheets}, which are designed to describe and document datasets. We use this benchmark generator to produce the benchmarks used in the evaluation. The end-to-end system, \sys{}, indexes content and context, accepts natural language questions and produces ranked tables as outputs.

\mypar{Evaluation Results} \sys{} surpasses state-of-the-art information retrieval systems, including full-text search with BM25~\cite{robertson2009probabilistic}, the RAG system implemented by LlamaIndex~\cite{Liu_LlamaIndex_2022}, and the advanced table retrieval system Solo~\cite{wang2023solo}, in retrieval quality and efficiency. \sys{} achieves up to a \textbf{22.95\% higher} relevant table hit rate compared to these baselines. \sys{} serves user queries up to \textbf{31$\times$ faster} than the baselines while requiring \textbf{orders of magnitude less storage} than baselines that produce numerous embeddings.

\mypar{Outline} We begin by introducing the preliminaries, including the definitions of context and content questions (Section~\ref{sec:datadiscovery}). We then provide an overview of \sys{}'s architecture (Section~\ref{sec:pneuma}), followed by a detailed explanation of its two key components: table representation (Section~\ref{sec:representation}) and table retrieval (Section~\ref{sec:pneuma_retriever}). In Section~\ref{sec:benchmark}, we describe the process of generating the table discovery benchmark, and in Section~\ref{sec:evaluation}, we present the evaluation results. Lastly, we present related work and conclusions.

\section{Content and Context Questions}
\label{sec:datadiscovery}

Data discovery is the problem of identifying and retrieving documents that satisfy an information need~\cite{aurum}. In the paper, we focus on tabular datasets: a table $D$ consists of a relational schema $\mathcal{R}(A_1, \dots, A_n)$ over $n$ columns, and a set of tuples $\mathcal{T}$, which are instances of the schema $\mathcal{R}$. Solving data discovery is challenging when there is a large volume of tables. For example, in large organizations, data is often distributed across multiple independently maintained databases. This creates information silos and requires extensive tribal knowledge to locate and access the relevant data. Additionally, many organizations use data lakes to store vast quantities of raw data, further complicating data discovery due to the sheer volume and lack of comprehensive metadata and documentation. Furthermore, scientific repositories, which archive experimental results and observational data, also present significant discovery challenges. Researchers need efficient methods to locate datasets relevant to their specific scientific questions.

While substantial research has explored discovering combinations of tables based on an initial table provided by the user~\cite{zhang2020web}, i.e., based on \emph{content}, there is significantly less work that identifies relevant tables based on their \emph{content} or \emph{context}. 

\mypar{Definitions} A table's content consists of column names and row values, which can be directly provided to a system. However, its context can only be generated by external agents, such as the person who collected the data or previously used the table for specific tasks. While table content is inherently structured, table context can range from free text to semi-structured documents. In this work, we treat contexts as free-form text that provides additional information about the table, which cannot be directly derived from its content.

\begin{figure}[ht]
    \centering
    \includegraphics[width=0.7\linewidth]{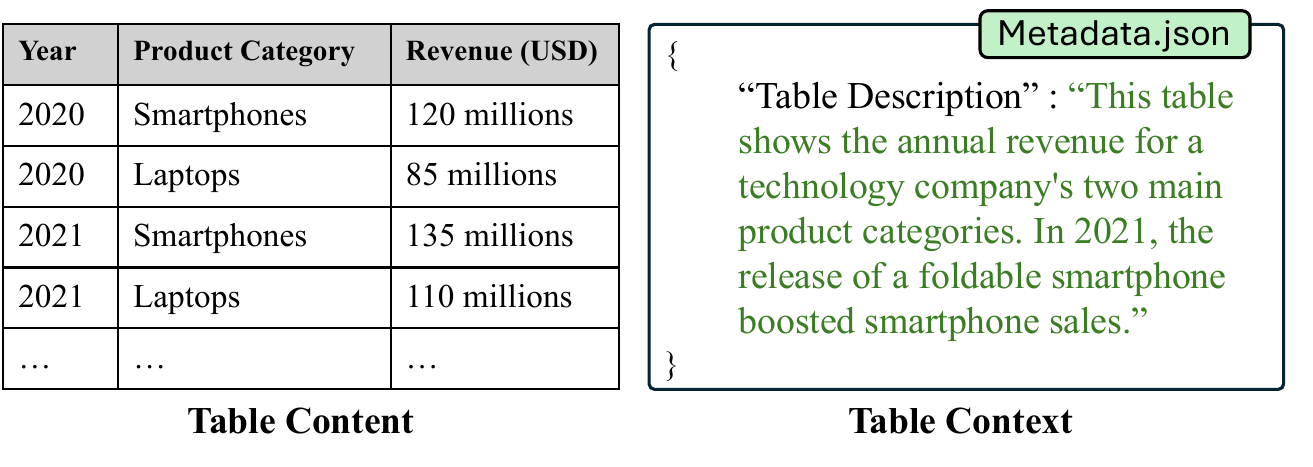}
    \caption{Example table content and context}
    \label{fig:example}
    \Description{Example table content and context}
\end{figure}

Figure~\ref{fig:example} presents an example of table content and context. On the left is a table showing the revenues of electronic products for a technology company. On the right is the table's context, the table description stored in a metadata file from an enterprise catalog.

A content question is answered by the table content, e.g., ``Which dataset contains information about annual revenue for smartphones and laptops?''. The table content is crucial for finding the table to answer this question, as it asks for specific entities within the table.

A context question is answered using the table's context rather than the table content.
The table context encompasses all information that informs the understanding and use of the data~\cite{DBLP:conf/cidr/HellersteinSGSA17}, such as metadata like column descriptions, how the data was generated, and its intended purpose. A crucial aspect of table context is that it cannot be inferred directly from the dataset. For instance, questions like "For what purpose was the dataset created?" and "What sampling method was used to generate this dataset?", which are important for practical data science~\cite{gebru2021datasheets}, can only be answered by external sources, such as the data creator. Datasheets for datasets~\cite{gebru2021datasheets} introduces a structured approach to eliciting dataset context through a series of questions aimed at data creators.
% \notera{can't we use the following examples to emphasize the reviewer who wanted to see more examples? e.g., M4??}
An example context question is ``Which dataset reflects the market impact of cutting-edge flexible display technology in the mobile industry?''. This question requires considering the table's context, as the context reveals that the company released a groundbreaking foldable smartphone in 2021.

\textbf{Where do data contexts come from?} This depends on the scenario. Sometimes, table contexts are readily available, such as in online repositories (ICPSR, Dataverse, or HuggingFace). HuggingFace' datasets page contains ``data cards'', which represent ``context''. In enterprise scenarios, table contexts are sometimes maintained in data catalogs. Additionally, alternative research lines provide approaches to generating such context, e.g., internal data markets to incentivize the creation of metadata/context~\cite{fernandez2020data}. \sys{} is designed to leverage data context when it is available. In the absence of data context, \sys{} can still perform table searches based solely on content.

\mypar{Today's Landscape and Problem Statement} Academic indexes, such as PubMed~\cite{pubmed} and JSTOR~\cite{jstor}, support table discovery via basic metadata such as titles, authors, or abstract descriptions. However, they often lack detailed data context, such as the dataset's purpose or the methods used for data collection. Additionally, these systems typically rely on keyword-based searches over metadata, offering a limited semantic understanding of user queries and search capabilities over the dataset content itself. Industry catalogs such as Google Dataset Search~\cite{brickley2019google} and Amundsen~\cite{amundsen}, face similar issues, relying heavily on complete and accurate metadata while neglecting the need to search through the data content and context. Internal resources, such as wikis or Slack channels, are often disorganized and reliant on tribal knowledge, further complicating access. 

The Retrieval-Augmented Generation (RAG) architecture~\cite{lewis2020retrieval} is a promising solution to the table discovery problem because it permits \emph{normalizing} the representation of content and context and searches over vectors. Furthermore, adding LLMs to the retrieval pipeline may improve ranking relevancy and, most importantly, enable querying tables using natural language.

\begin{problem}[Table Discovery]
Given a natural language question $Q$, a collection of tables $\mathcal{D}$, a corpora of table context $X_{\mathcal{D}}$ and a table discovery system $S$, find the relevant table $D_Q \in \mathcal{D}$ that is relevant to answer $Q$, i.e., $D_Q = S(Q, \mathcal{D}, \mathcal{X}_{\mathcal{D}})$. 
\end{problem}

\section{\sys{} Architecture}
\label{sec:pneuma}

\begin{figure}[ht]
    \centering
    \includegraphics[width=0.7\linewidth]{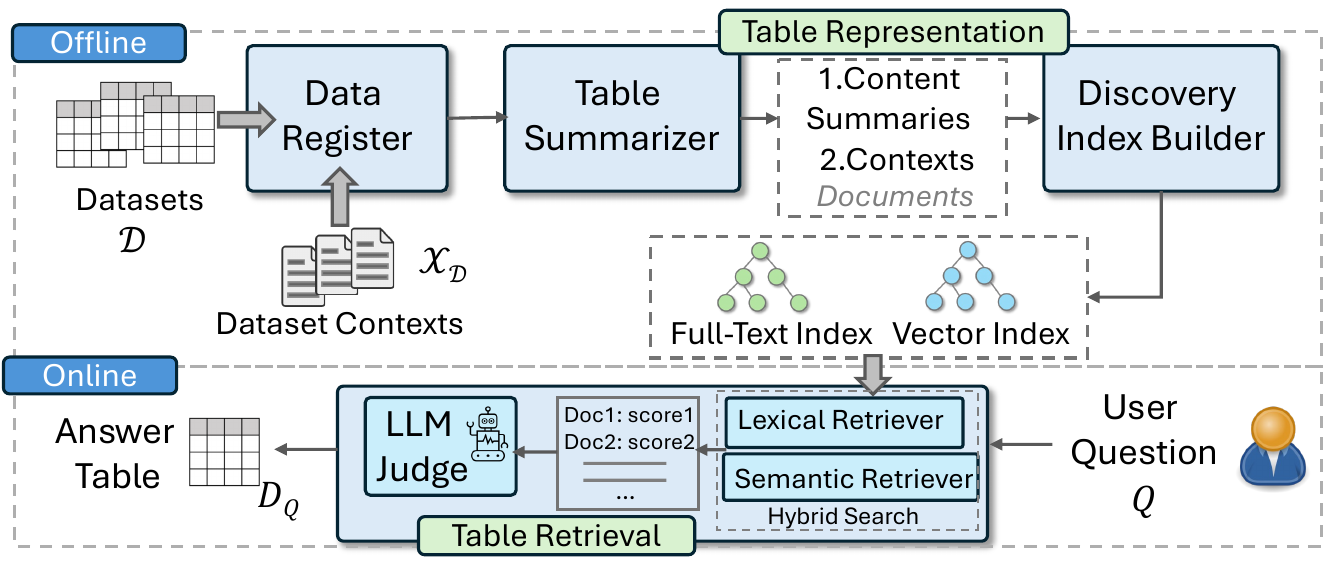}
    \caption{System Overview of \sys{}}
    \label{fig:pneuma_archi}
\end{figure}

In this section, we present the design of \sys{} and explain how it addresses the table discovery problem using content and context. \sys{}'s offline stage focuses on table representation and its online stage on table retrieval. Figure~\ref{fig:pneuma_archi} illustrates the system architecture of \sys{}.

During the offline phase, users register tables and associated contexts via the \textsf{Data Register} component, which offers APIs to represent the context and to ingest tables from different sources such as databases and CSV files. The volume of table collections can be large, thus \sys{} employs the \textsf{Content Summarizer} to represent large tables into smaller documents, referred to as ``content summaries'', while preserving schema and row-level information. These summaries, along with table contexts, are indexed by the \textsf{Discovery Index Builder} into both full-text and vector indices, enabling efficient table retrieval. At this stage, \sys{} treats both data content and context uniformly as text documents.

During the online phase, \sys{} retrieves tables based on the user query $Q$ by integrating three signals: lexical (BM25), semantic (vector search), and a signal based on LLM judgment. The lexical and semantic retrievers complement each other—BM25 handles exact lexical matches, while vector search captures semantic similarity, even without exact matches. To harness both strengths, \sys{} scores document relevance using both retrievers. These scored candidates are then passed to \textsf{LLM Judge}, which determines whether each candidate document is relevant to the query. Based on these judgments, \sys{} reorders the documents, and returns the top-ranked tables associated with the retrieved documents.

\noindent\textbf{Support Content and Context Questions.} 
\sys{} seamlessly supports table discovery queries based on table content, context or both via its two core mechanisms: \textit{table representation} (offline) and \textit{table retrieval} (online). \textsc{Pneuma} is designed to leverage table context when it is available but table contexts need not be available for \textsc{Pneuma} to work. Users can point \sys{} to a folder containing tables and their associated contexts, which are then transformed into normalized representations and indexed. Once the indexing process is complete, users can query the system using natural language questions. The table retrieval mechanism ensures that relevant documents—whether content or context—are retrieved in response to user queries. For example, content-based questions like "What were the total sales for a specific product in the last quarter?" are answered by retrieving relevant table content summaries. Similarly, context-based questions, such as "Find datasets created using randomized controlled trials to evaluate the effectiveness of new treatments", are addressed by retrieving the appropriate context documents. This approach eliminates the need for manual search and allows users to discover relevant tables simply by asking questions in natural language.

The next two sections introduce table representation and table retrieval in more detail.

\section{Table Representation}
\label{sec:representation}

Effectively representing tables is essential for accurate and efficient table retrieval. Some approaches~\cite{herzig2021open, yin2020tabert} represent tables by embedding them into vector spaces, typically trained for specific tasks like table question-answering~\cite{jin2022survey}. However, these approaches often lose row-level details, which are critical for precise data retrieval, especially for queries targeting specific facts or values. Additionally, these methods are resource-intensive to train and do not generalize well across new table corpora or tasks~\cite{wang2023solo}. Other approaches~\cite{Liu_LlamaIndex_2022, wang2023solo} split tables into smaller text snippets, typically at the row level, treating each snippet as a document. While this preserves row-level information, it leads to scalability issues, as storage requirements grow linearly with the number of rows, creating significant overhead. This approach also slows down the indexing process for large datasets containing millions of rows. For instance, \rag~\cite{Liu_LlamaIndex_2022}, a popular RAG library, produces a 44 GB index for the Chicago Open Data (829 MB) and a 26 GB index for a subset---500 rows max per table---of BIRD~\cite{BirdBenchmark} (218 MB). These indexes are much larger than the original datasets and hence harder to justify their storage costs. On the other hand, \sys{} produces 458 MB and 23.9 MB indexes for the Chicago Open Data and the BIRD dataset, respectively.

Table-level embedding approaches struggle with questions that require keyword matching at the row level, as the raw text information often gets lost in the embedding process. On the other hand, indexing every row in a table leads to high storage overhead and neglects the schema’s structural context. Indexing only the table schema might save space, but it becomes ineffective when many tables share similar schemas, making it hard to differentiate between them. Moreover, real-world datasets often contain ambiguous or uninformative column names, so merely concatenating column names to represent the schema is not sufficient. Thus, a well-rounded table representation must integrate both schema and row-level information while retaining some raw textual information.

To address these challenges, \sys creates table representations by generating both schema summaries and a smaller number of row summaries. Schema summaries describe the table's columns, preserving structural information crucial for understanding the relational context of the data. Row summaries, on the other hand, combine row values, providing a detailed representation of the table's content. 
\sys{}'s \textsf{Table Summarizer} leverages an LLM to generate schema summaries. Additionally, \sys{} optimizes the generation process by dynamically adjusting the inference batch size, filling each batch size with tasks of similar computational workload for balanced and efficient processing. We next introduce how \sys generates schema and row summaries.

\mypar{Generate schema summaries} Concatenating column names alone is insufficient for effective schema summaries, as column names in many tables are often cryptic or abbreviated. For example, the columns "AB" and "BABIP" are unclear. To address this, \sys{} uses an LLM to generate schema summaries by providing a meaningful description of each column. Here, the LLM recognizes the "AB" column as the number of at-bats and the column "BABIP" as an abbreviation of Batting Average on Balls In Play, then narrates each with more descriptions. These narrations are then concatenated to form schema summaries, as illustrated in Figure~\ref{fig:illust_schema_summaries}. To ensure accuracy and coherence, \sys{} provides the full schema as context to the LLM when prompting it to describe individual columns.

\mypar{Generate row summaries} Row summaries provide more detailed information about the table content and differentiate tables with similar schemas. For instance, in our evaluation using the FetaQA~\cite{fetaqa} where many tables share the same schema, row information was critical for improving the hit rate. To address this, \textsc{Pneuma} randomly samples $r$ rows, as it avoids making any assumptions about the table content. For each sampled row, \textsc{Pneuma} concatenates its values with their respective column names to generate $r$ summaries, as illustrated in Figure~\ref{fig:illust_row_summaries}. Including column names is essential, as it provides context for each value, ensuring clarity in the data's meaning. Without the column names, values like "76" could be ambiguous, potentially representing a variety of attributes, such as age, quantity, or price. By pairing values with their respective columns, \sys{} maintains clarity and structure, allowing for more accurate and contextually aware table retrieval.

\begin{figure}[ht]
    \centering
    \begin{minipage}{0.6\textwidth}
        \centering
        \fbox{
            \begin{minipage}{0.9\textwidth}
                {\scriptsize \ttfamily
                AB: This represents the number of at-bats for a batter. |
                BABIP: This represents Batting Average on Balls In Play, a statistic that measures a batter's success rate in converting balls put into play into hits. | ...
                }
            \end{minipage}
        }
        \caption{An example schema summary}
        \label{fig:illust_schema_summaries}
    \end{minipage}%
    \hfill
    \begin{minipage}{0.4\textwidth}
        \centering
        \fbox{
            \begin{minipage}{0.9\textwidth}
                {\scriptsize \ttfamily
                AB: 76 | BABIP: 0.317 | ...\\
                AB: 11 | BABIP: 0.111 | ...\\
                ...
                }
            \end{minipage}
        }
        \caption{An example row summary}
        \label{fig:illust_row_summaries}
    \end{minipage}
\end{figure}

\mypar{Data ingestion via dynamic batch-size selection} \sys{} employs an LLM to generate schema summaries for each column in a table collection, requiring one inference per column. For large collections, this leads to a large number of LLM calls, making LLM inference the primary computational bottleneck during the ingestion process. Thus, optimizing LLM inference time is crucial for the practical scalability of \sys{} on large datasets. For instance, on the FeTaQA dataset, which contains over 10K tables, LLM inference without optimization requires approximately 15 hours. With \sys{}'s optimization techniques, this time is reduced to just 2 hours. We detail how \sys{} achieves this improvement below. 

Batch processing offers a way to optimize LLM inference by allowing multiple inputs to be processed in parallel during a single forward pass through the LLM~\cite{agrawal2024tamingthroughputlatencytradeoffllm}. By grouping several requests into a batch, the system leverages parallel computation on GPUs, reducing processing time and improving resource utilization. However, determining the optimal batch size for LLM inference is not straightforward. We cannot simply choose a batch size based on available GPU memory because the memory usage varies depending on the complexity and token length of the prompt. Larger batches improve throughput but demand more memory, potentially causing out-of-memory errors, while smaller batches underutilize resources and reduce efficiency. To address this challenge, \sys{} dynamically adjusts batch size during inference, automatically determining the optimal batch size for efficient processing.

\sys{} begins by pooling prompts, where each prompt instructs the LLM to generate a schema description for a table column. It then employs a binary search algorithm to identify the optimal batch size within a predefined range. If a batch size exceeds memory limits and triggers an out-of-memory error, \sys{} reduces the batch size; conversely, if a batch size runs without errors for several iterations, \sys{} increases it. To further optimize memory usage, \sys{} sorts the prompts by size and assigns them to each batch in a way that ensures each batch contains a similar number of tokens, balancing memory requirements across batches for maximum efficiency. For example, when processing the Adventure Works dataset, we get a batch size of 50 instead of 14 using this balancing approach. For reference, this reduces the summarization time by 78\% (from 288 to 63 seconds).

\section{Table Retrieval}
\label{sec:pneuma_retriever}

We present the table retrieval method in Section~\ref{subsec:retrieval} and indexes in Section~\ref{subsec:indexbuilder}.

\subsection{\sys{}'s Hybrid Retrieval}
\label{subsec:retrieval}

In the online stage, \sys{} retrieves a ranked list of tables for a user's query by integrating three signals: lexical (BM25~\cite{robertson2009probabilistic}), semantic (vector search over embeddings of context and content summaries), and the LLM judgment signal. These signals are combined to ensure robust and accurate retrieval across a wide range of user queries. First, lexical and semantic retrieval methods are used to generate candidate tables and their initial relevance scores. The LLM then evaluates these candidates to refine the relevance ranking. While traditional retrieval methods like BM25 and vector search are effective for identifying relevant documents, they may miss subtle nuances in the query or context. The LLM receives a ranked list of candidate tables provided by the two retrievers, and refines the results for better precision, especially in complex or ambiguous queries where context and meaning are critical.

The novelty of \sys{}'s approach lies not in any single signal, but in the integration of all signals. During table retrieval, \sys{} grounds the LLM with factual information—relevance scores—from two reliable retrievers. Unlike many approaches that rely on the LLM for re-ranking, our method assigns the LLM a simpler, well-defined task: to judge whether a candidate document (content/context) is relevant to the question. In other words, we decompose an otherwise difficult task for the LLM, improving its performance.

\noindent\textbf{Hybrid Search.} \textsc{Pneuma} adopts a hybrid search strategy that utilizes both full-text and vector retrievers to retrieve and rank relevant documents. \sys{} retrieves $n*k$ documents independently from both the full-text and vector retrievers. Here, $k$ represents the number of documents the user wants to inspect, which is explicitly set by a user. Meanwhile, $n$ is a multiplicative factor that determines the number of documents to retrieve initially for each retriever. $n$ is transparent to the user. $n$ ensures that the \textsf{LLM Judge} has a larger pool of candidates to evaluate.

Next, each unique document, from a set of up to $2nk$ documents, is assigned a combined relevance score. This score is calculated by weighting the relevance scores from both retrievers. If a document is retrieved by only one retriever, its relevance score for the other retriever is also computed. The final combined score for a document, $s(d)$, is evaluated as follows:
\[
s(d) = \alpha \cdot s_{\text{lexical}}(d) + (1 - \alpha) \cdot s_{\text{semantic}}(d)
\]

Where $s_{\text{lexical}}(d)$ is the normalized relevance score (using min-max scaling) from the full-text retriever, $s_{\text{semantic}}(d)$ is the normalized score from the vector retriever, $\alpha$ is a weighting factor (between 0 and 1) that balances the influence of both retrievers. 

This scoring function allows \sys{} to integrate the strengths of both retrieval methods, ensuring a more accurate ranking of documents based on both lexical and semantic relevance.

\mypar{LLM Judge} The performance of table retrieval can be further improved by \textsf{LLM Judge}. Formally, its task is to classify pairs of documents and questions as either relevant or not. We reorder the documents based on the LLM’s output. As we iterate through the list, any document marked as irrelevant by the LLM is moved to the end, ensuring the most relevant documents are prioritized at the top. At the same time, the relative order between documents deemed relevant by the LLM is preserved based on their scores from the previous step.

The output of \textsf{LLM-Judge} is the final list of relevant documents to a user's question. The tables associated with the retrieved documents are then returned to the user.

\noindent\textbf{Illustration of \sys{}'s Retrieval Mechanism} We use a real example to demonstrate the benefits of \sys's retrieval mechanism. Consider this context question from ChEMBL (a chemical database) \textit{``Develop a comprehensive database that catalogs and organizes the fundamental characteristics and attributes of tiny organic substances, encompassing their structural compositions, mass, and other relevant details.''}. The correct table context to address this question is \textit{``The dataset was created to provide a comprehensive repository of chemical compounds, specifically small molecules, with their associated properties and information. The primary task in mind was to fill a gap in the availability of structured data on approved and investigational small molecule drugs, including their chemical structures, therapeutic uses, and regulatory information.''}

At $k=1$, neither the full-text or vector-based retriever retrieved this document or its associated table. The full-text retriever failed due to limitations in lexical matching, while the vector-based retriever struggled because the pre-trained embedding model does not fully capture the nuanced semantics in specialized domains like chemistry. In contrast, \sys's hybrid search successfully retrieved this context using its vector retriever at a lower rank. It then elevated the context to the top rank by weighting the combined score. \textsf{LLM Judge} validated the relevance of this context to the question, keeping it in the first rank. Consequently, Hybrid Search + LLM Judge correctly retrieved the relevant context and associated table for the question.

In summary, \sys{} achieves robust table retrieval by leveraging the complementary strengths of lexical and semantic retrieval, combined with the advanced semantic understanding of LLMs, to handle a wide range of user queries effectively.

 \subsection{Efficient Index Construction}
 \label{subsec:indexbuilder}

To support the table retrieval in \sys{}, we need to build efficient data discovery indices during the offline stage. \sys{} needs to index two types of documents: content summaries and table contexts. Unlike content summaries, table contexts are already in text form and can be treated directly as documents without passing through the \textsf{Table Summarizer}. At this stage, \sys{} treats both content and context uniformly, viewing them as documents for indexing.

To support table retrieval, \sys{} builds two types of indices: full-text and vector indices. Each of these indices processes all documents, associating them with their respective table and document type—either as a content summary or a table context. 

\mypar{Block-based Embedding Generation} Building the vector index requires generating embeddings for documents in advance. The challenge is that in \sys{}, each document is small, and there are a large number of documents to process, making the indexing process time-consuming and storage-intensive. To improve the indexing time and storage overhead, we group documents from the same table and of the same type into fixed-size blocks determined by the context window size of the encoding model. For instance, if there are three row summaries for Table A, rather than embedding each summary individually, we combine them into a single document and then generate a single embedding for the block. This reduces the number of embeddings required, accelerates the indexing process, and lowers storage overhead, while still preserving all relevant information for retrieval. It also avoids scenarios where a given document is larger than an LLM's context window because the document size is bounded by the window size of the embedding model which is typically much smaller than the LLM's context window.

\noindent\textbf{Incrementally Maintaining the Index.} When there is an update to the schema, \textsc{Pneuma} can generate new schema summaries for the updated data by calling \textsf{Table Summarizer} and then insert the new summaries into the index. For row updates, one could similarly call \textsf{Table Summarizer} to do the re-sampling, either eagerly, or only after a defined threshold is exceeded, e.g., when more than 20\% of the rows are updated. \textsc{Pneuma} does not have a built-in mechanism to detect those changes because it makes no assumptions of the storage system that keeps those tables. For instance, they may be stored in S3. But \textsc{Pneuma} is engineered to expose an API for updating that can be invoked by external clients.

\section{Table Discovery Benchmark}
\label{sec:benchmark}

Existing data discovery benchmarks are limited. NQ\_tables~\cite{herzig2021open}, FeTaQA~\cite{fetaqa}, and QATCH~\cite{QatchBenchmark}, which are designed for table question answering, often reveal the correct answer table in the question itself. CMDBench~\cite{CMDBench} targets table discovery but suffers from high false-negative rates due to incomplete ground truth. Additionally, BIRD~\cite{BirdBenchmark} and Spider~\cite{yu2018spider}, which were developed for text-to-SQL tasks, focus on table content and overlook table context, limiting their ability to evaluate systems that leverage content and context.
% \notera{we are saying BIRD is not useful but then we use it in the evaluation?} \noteltf{I agree. I think we need to perhaps exclude BIRD here or add extra texts saying it is still useful as it provides content questions.}

To address this gap, we introduce a benchmark generator that produces benchmarks specifically designed for table discovery, covering both content and context questions. By incorporating both types of queries, a benchmark simulates a diverse range of real-world scenarios, allowing for a more comprehensive evaluation of a system's capabilities. 
% 
% We consider five datasets for our experiments: ChEMBL~\cite{chembl}, a scientific database focused on bioactivity data of drug-like molecules; Adventure Works~\cite{AdventureWorks}, a database simulating an enterprise environment for business operations; Chicago Open Data~\cite{chicagodataportal}, a repository containing public data from the City of Chicago; PublicBI~\cite{PublicBIBenchmark}, a public business intelligence dataset, and FeTaQA~\cite{fetaqa}, a dataset designed for factual table question-answering tasks. For each dataset, we generate 2K questions, half focusing on content and half on context.
% 
A benchmark consists of two components: the content benchmark and the context benchmark. Each benchmark maps a question to a list of tables, where each individual table within the list can answer the question. The content benchmark focuses on content questions, represented as $(Q_c, [T])$, while the context benchmark addresses context questions, represented as $(Q_x, [T])$.
% 
% focusing on content-related and context-related questions, respectively. Each benchmark comprises pairs $(Q, \mathcal{T})$, where $Q$ is a question, and $\mathcal{T}$ is a set of tables that can answer the question.
% 
In the following sections, we detail the process of generating both content and context benchmarks.

\subsection{Content Benchmark}

For the content benchmark, our goal is to generate questions that can be directly answered by the content of a table. A straightforward approach is to prompt powerful LLMs, such as GPT-4~\cite{achiam2023gpt}, to generate questions based on several sample rows from a table. However, this approach is insufficient for two reasons. First, LLMs are prone to hallucination, often generating overly simplistic or irrelevant questions that may not reflect the complexity of the data. Second, there is little control over the diversity of the questions generated—such as ensuring they cover different aspects of the table (e.g., addressing various columns)—which is critical for evaluating table discovery systems comprehensively.

Therefore, we use SQL queries as an intermediate proxy for generating natural language questions. This approach complements LLMs by leveraging the structured nature of SQL to ensure both relevance and diversity. SQL queries inherently target specific columns and operations, controlling the scope of questions and reducing the likelihood of hallucinations. Given a table, we first generate SQL queries based on the table’s structure. These SQL queries are then converted into natural language questions. We introduce these two stages below.

\subsubsection{SQL Generation} Recent work, \textsc{Solo}~\cite{wang2023solo}, has demonstrated that using a SQL template is an effective approach for generating meaningful questions. While \textsc{Solo} follows the simple template defined in WIKISQL~\cite{zhong2017seq2sql}, we extend it to include more complex SQL operators such as \textsf{Group By}, \textsf{Having} and \textsf{Order By}. In our template, which is shown in Figure~\ref{fig:sql_template}, we exclue \textsc{Join} operations, focusing on questions that can be answered by a single table. Additionally, the \textsc{From} clause is omitted because identifying the table identifier is the goal of table discovery systems, and thus explicitly including the table name is forbidden in question generation. By omitting the \textsc{From} clause, we guarantee that the table name will not appear in the generated question.

We generate SQL queries by filling in the SQL template through random sampling. Specifically, we randomly select columns and corresponding values within the selected columns. Additionally, we assign probabilities for including aggregation functions, \textsc{Group By}, \textsc{Having}, and \textsc{Order By} clauses, and include them based on those probabilities. We ensure that the generated SQL queries maintain valid syntax even if certain clauses are not selected (with the exception of the \textsc{From} clause, which is omitted by design).

\begin{figure}[ht]
    \centering
    \footnotesize
    \begin{minipage}{0.8\linewidth}
        \centering
        \begin{lstlisting}[style=sqlStyle]
            SELECT column_1, MAX(column_2)
            WHERE column_3 = value_1 AND column_4 < value_2
            GROUP BY column_1 HAVING MAX(column_2) > value_3
            ORDER BY column_1 LIMIT 3;
        \end{lstlisting}
    \end{minipage}
    \caption{Example SQL Query Template}
    \label{fig:sql_template}
\end{figure}

\subsubsection{Question Generation} Next, we use an LLM to convert SQL into a natural language question. The LLM receives the table schema, a few sampled rows, and the SQL query. However, the initial questions often suffer from quality issues, such as overusing keywords from cell values or inconsistencies due to LLM hallucinations. We address these issues and enhance question quality through the following two stages.

\mypar{Question Rephrasing} To better simulate real-world user queries, we ensure LLM-generated questions do not directly replicate column names or long cell values, as users often cannot recall exact details in practice. To implement this, we identify questions that contain long cell texts matching the associated table data. We then instruct the LLM to paraphrase the questions by rewording them to avoids exact matches while maintaining original meaning.

\mypar{Consistency Check} Even when the LLM is guided to generate questions based on SQL queries, it may still produce hallucinations due to its inherent limitations. This can result in questions that are inconsistent with the original SQL query. For example, the SQL may be designed to return two columns, but the generated question might only reference one of them. Additionally, the LLM can confuse column names that look similar, leading to inaccuracies in the question. To ensure consistency between the SQL query and the generated question, we apply the concept of \textit{Cycle Consistency}, originally introduced in machine translation \cite{lample2017unsupervised}. Specifically, we ask the LLM to translate the generated question back into SQL (referred to as the Back SQL) and compare it with the original SQL. We use the SQL exact match metric from \cite{yu2018spider} to evaluate SQL equivalence, which requires the two SQL queries to have the exact same components. If the Back SQL does not match the original SQL exactly, we consider the question inconsistent and discard it.

\subsubsection{Annotating other answer tables}
\label{sec:content_other_answer}

There may be more than one table capable of answering a given question. This introduces a challenge when designing a benchmark, as it could lead to false negatives.
% 
% The table of which a question is generated is considered the answer table. However, other tables may also be capable of answering the same question. For example, if another table is identical to the initial answer table, it could serve the same purpose. Ensuring that a question has only one unique answer table is difficult. This presents a challenge for benchmarks, as it could lead to false negatives. 
Specifically, a system might identify a valid table that answers the question, but it could be unfairly penalized for not retrieving the specific table on which the question was originally generated. To address this, we extend our search to identify other tables that can also answer the generated question. Each question corresponds to a SQL query during content question generation. Thus, a table is identified as an alternative answer table for a question if it contains all the columns and values referenced in the corresponding SQL query of the question.

\subsection{Context Benchmark}

Context questions are those that can only be answered by considering the broader context of a table, rather than its content alone. Therefore, generating context questions requires first establishing the contexts associated with each table.

\subsubsection{Context Generation}

A table's context is typically provided by external agents, such as the data creator or individuals familiar with its use. However, manually creating contexts for our table collection, which includes 11,501 tables in total, would be prohibitively expensive. To overcome this issue, we leverage large language models (LLMs), which have shown significant potential in role-playing scenarios~\cite{shao2023characterllmtrainableagentroleplaying}. We instruct an LLM to assume the role of the table creator and automatically generate context for each table.

But how do we prompt the LLM to elicit relevant table context? Datasheets for Datasets~\cite{gebru2021datasheets} propose 51 questions aimed at data creators to gather comprehensive data contexts. These questions cover topics such as the reasons for creating the dataset, data collection methods, and usage considerations.

For each table, we prompt the LLM (acting as the data creator) to answer this set of questions, thereby generating detailed context automatically. 
In the prompt, we emphasize two key characteristics—completeness and relevance—to ensure the LLM produces high-quality responses. Completeness means that the answer definitively and comprehensively addresses all parts of the question, particularly important for long and compounded questions. Relevance focuses on providing only the information requested, suppressing the LLM’s tendency to include extraneous details. These criteria are crucial for generating high-quality, focused answers that provide the necessary context for each table while avoiding unnecessary or off-topic information. For reference, the LLM we use for context generation, Meta-Llama-3-8B-Instruct, takes approximately 4 minutes to generate contexts for each table (batch size $k=1$), which are essentially answers to the 51 context-elicitation questions.

\subsubsection{Context Benchmark Generation} Next, we generate the context benchmark by prompting an LLM to create questions based on table contexts. We use stratified sampling to select 1,020 contexts, with 20 contexts for each of the 51 context-elicitation questions defined in \cite{gebru2021datasheets}. For each selected context, we prompt the LLM to generate a question that asks for a table based on that context. The table associated with the context becomes the answer table for the generated question. However, many of the generated questions contain keywords directly taken from the context, which does not accurately reflect real-world user queries. To address this, we have the LLM rephrase the questions to make them more natural and realistic, which is similar to the process we follow for content.

\subsubsection{Annotate other answer tables}
\label{subsubsec:label_bx1}

Other tables may have relevant contexts that can also answer a question in the context benchmark.
Similar to what we explained in Section~\ref{sec:content_other_answer}, it is important to consider these tables to avoid unfairly penalizing a table discovery system during evaluation. To identify additional answer tables, we prompt an LLM to check which other contexts are capable of answering a given question.

\section{Evaluation}
\label{sec:evaluation}
In this section, we answer the following research questions:

\begin{myitemize}
    \item \textbf{RQ1 (End-to-end evaluation):} Does \sys{} effectively identify the relevant tables for the given questions?
    \item \textbf{RQ2 (Efficiency and Scalability):} Is \sys{} efficient and scalable in answering user questions during the online stage? Additionally, does it optimize storage footprint and data preparation time in the offline stage?
    \item \textbf{RQ3 (Ablation study):} Is each component within \sys{} useful? Is the design of each \sys{} component well-justified?
    \item \textbf{RQ4 (Microbenchmarks):} How do \sys{}'s hyperparameters affect its performance?  We also investigate whether \sys{} is sensitive to the choice of LLMs and evaluate its performance on a simpler benchmark where the questions have significant keyword overlap with the data.
\end{myitemize}

\mypar{Datasets} We evaluate \sys using six real-world datasets (Table~\ref{tab:datasets}), spanning public, scientific, business, and enterprise data environments. This variety ensures comprehensive testing across different workloads and scenarios. Specifically, we apply our content and context benchmark generators to ChEMBL, Adventure Works, Public BI, Chicago Open Data, and FeTaQA, resulting in 5 content benchmarks and 5 context benchmarks. In addition, we leverage an external benchmark BIRD~\cite{BirdBenchmark}, which provides content questions. Table~\ref{tab:datasets} shows the statistics of these datasets.

\begin{table}[ht]
\caption{Datasets Characteristics}
  \centering
  \begin{tabular}{||c|c|c|c|c||}
    \hline
    \textbf{Name} & \textbf{\#Tables} & \textbf{Avg. \#Rows} & \textbf{Avg. \#Attributes} & \textbf{Size} \\
    \hline\hline
    ChEMBL & 78 & 5,161 & 7 & 67 MB \\
    \hline
    Adventure Works & 88 & 9,127 & 8 & 102 MB \\
    \hline
    Public BI  & 203 & 20 & 66 & 2.2 MB \\
    \hline
    Chicago Open Data & 802 & 2,812 & 17 & 829 MB \\
    \hline
    FeTaQA & 10,330 & 14 & 6 & 42 MB \\
    \hline
    BIRD & 597 & 614,450 & 7 & 17 GB \\
    \hline
  \end{tabular}
  \label{tab:datasets}
\end{table}

\begin{myitemize}
    \item \textbf{ChEMBL}~\cite{chembl}: a scientific database containing bioactivity data on drug-like molecules, used in pharmaceutical research.

    \item \textbf{Adventure Works}~\cite{AdventureWorks}: simulates an enterprise database, with tables related to business operations, sales, and manufacturing. 

    \item \textbf{Public BI}~\cite{PublicBIBenchmark}: datasets used for business intelligence and data warehousing tasks. It contains data typically found in reports, dashboards, and business decision-making systems.
    
    \item \textbf{Chicago Open Data}~\cite{chicagodataportal}: contains publicly available data from the City of Chicago, representing various aspects of civic operations, including transportation, crime, and public services.

    \item \textbf{FeTaQA}~\cite{fetaqa}: a dataset designed for factual table question answering (QA) tasks, where users pose questions to retrieve factual information from tables.

    \item \textbf{BIRD}~\cite{BirdBenchmark}: a cross-domain dataset designed for evaluating text-to-SQL tasks, covering more than 37 professional domains, such as healthcare and blockchain.
\end{myitemize}

\mypar{\sys{} Implementation} We build \sys{} using Python 3.12.2. The LLM inside \textsf{Table Summarizer} and \textsf{LLM Judge} is Qwen2.5-7B-Instruct~\cite{Qwen2024}.\footnote{\url{https://huggingface.co/Qwen/Qwen2.5-7B-Instruct}} To generate row summaries, we set $r=5$, meaning we sample five rows per table. The lexical retriever is implemented using BM25s~\cite{BM25s}, a leading Python implementation of BM25. For the vector index, we employ a SOTA vector database, ChromaDB~\cite{chromadb}, which uses the HNSW index~\cite{malkov2018efficient} to support efficient vector search. 

\mypar{System Setup} All experiments are conducted on a DGX A100 server, which has 1 TB RAM and dual AMD EPYC 7742 CPUs (128 cores in total). In terms of software, this server has Python 3.12.2, CUDA 12.4, and Nvidia driver 550.90.07. For RQ1 and RQ2, we use these hyperparameters for \sys{}: $\alpha = 0.5$ and $n = 5$. We study the sensitivity of these hyperparameters in Section~\ref{subsec:microbenchmarks}.

\mypar{Baselines}
We compare \sys{} against the following baselines, which are state-of-the-art systems for table discovery.

\begin{myitemize}
\item \textbf{\fts{}}: Treats tables as plain text documents and builds a full-text search index over them. It then applies the BM25 algorithm to retrieve tables that best match a user's query. We implement this using a state-of-the-art BM25 library~\cite{BM25s}.

\item \textsc{\textbf{LlamaIndex}}: We implement a RAG system using LlamaIndex~\cite{Liu_LlamaIndex_2022}, a widely-used open-source library for retrieval-augmented generation. The system utilizes the \textsc{DBReader} API provided by LlamaIndex, which converts each row in a table into a vector representation, allowing the LLM to retrieve relevant table information based on user queries. 

\item \textsc{\textbf{Solo}}: Solo~\cite{wang2023solo} is a state-of-the-art system for table retrieval that employs a self-supervised learning approach. Given natural-language questions, it returns the most relevant table as output.

\end{myitemize}

\subsection{RQ1: End-to-End Evaluation}
\label{subsec:quality}

We use \textbf{hit rate} to evaluate \sys{}'s performance in identifying tables relevant to the given questions. In our benchmarks, while a question may have multiple relevant tables, any one of these tables is sufficient to answer the question fully. Therefore, the hit rate is defined as the proportion of questions for which \textit{at least} one relevant table appears in the top-$k$ retrieved documents. Next, we show the overall hit rates across context and content benchmarks, and we break down the hit rates over each benchmark.

\subsubsection{\sys{}'s Overall Hit Rate} 
\begin{figure}
    \begin{minipage}[t]{0.6\linewidth}
        \centering
        \subfloat[\centering Overall Hit Rates ($k=1$)]{{\includegraphics[width=\linewidth]{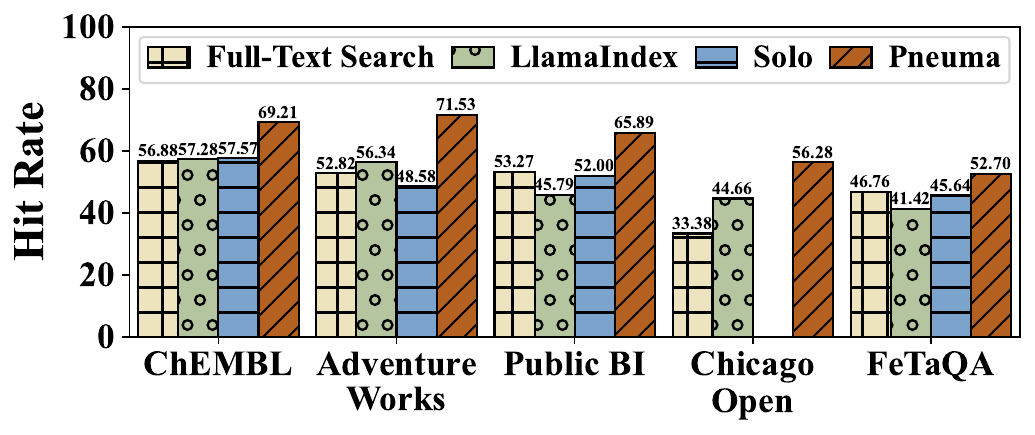} }}%
    \end{minipage}
    \begin{minipage}[t]{0.388\linewidth}
        \centering
        \subfloat[\centering Overall Hit Rates ($k=5$)]{{\includegraphics[width=\linewidth]{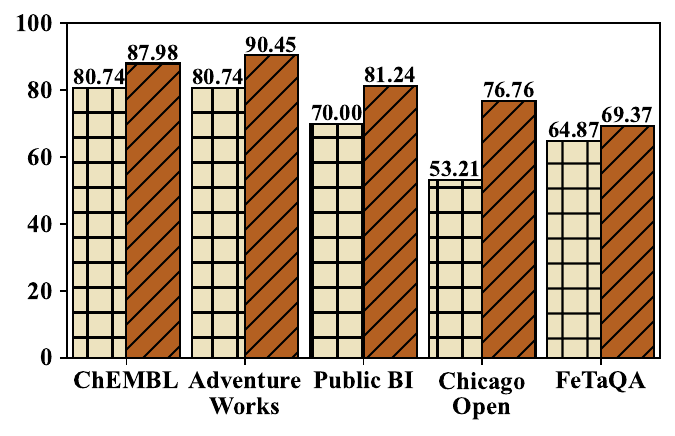} }}%
    \end{minipage}
    \caption{Overall Hit Rates on Context and Content Benchmarks ($k \in \{1,5\}$)}
    \label{fig:rq1_overall}
\end{figure}

Figure~\ref{fig:rq1_overall} shows the overall hit rate, over both content and context questions (2k questions in total), of \sys{} and the baselines at $k \in \{1,5\}$. When $k=1$, all systems must rank the relevant documents, associated with one of the correct answer tables, at the top. On the other hand, $k=5$ represents a scenario where users want to inspect more documents.

At $k=1$, \sys{} outperforms all baselines in every scenario. For example, on the Adventure Works variant, \sys{} obtains a hit rate of 71.53\%, which is 18.71, 15.19, and 22.95 percentage points more than \fts, \rag, and \solo, respectively. \solo cannot be run on the Chicago dataset due to its excessive vector generation (730 million vectors, $\sim$1 TB storage). Additionally, \solo is excluded at $k=5$ because its hyperparameter $k$ refers to the number of tables, making it incomparable to other baselines. \rag is also excluded at $k=5$ due to its overly slow performance at $k=1$.

At $k=5$, \sys{} still outperforms \fts, but the difference is smaller compared to $k=1$. This is because as $k$ increases, all approaches can include more tables and achieve higher hit rates. Nevertheless, it is still considerable because even a $10\%$ difference corresponds to approximately 202 questions.

\subsubsection{\sys{}'s Hit Rate in Answering Content Questions}
\begin{figure}
    \centering
    \subfloat[\centering Hit Rates on Content Benchmarks]{{\includegraphics[width=0.75\linewidth]{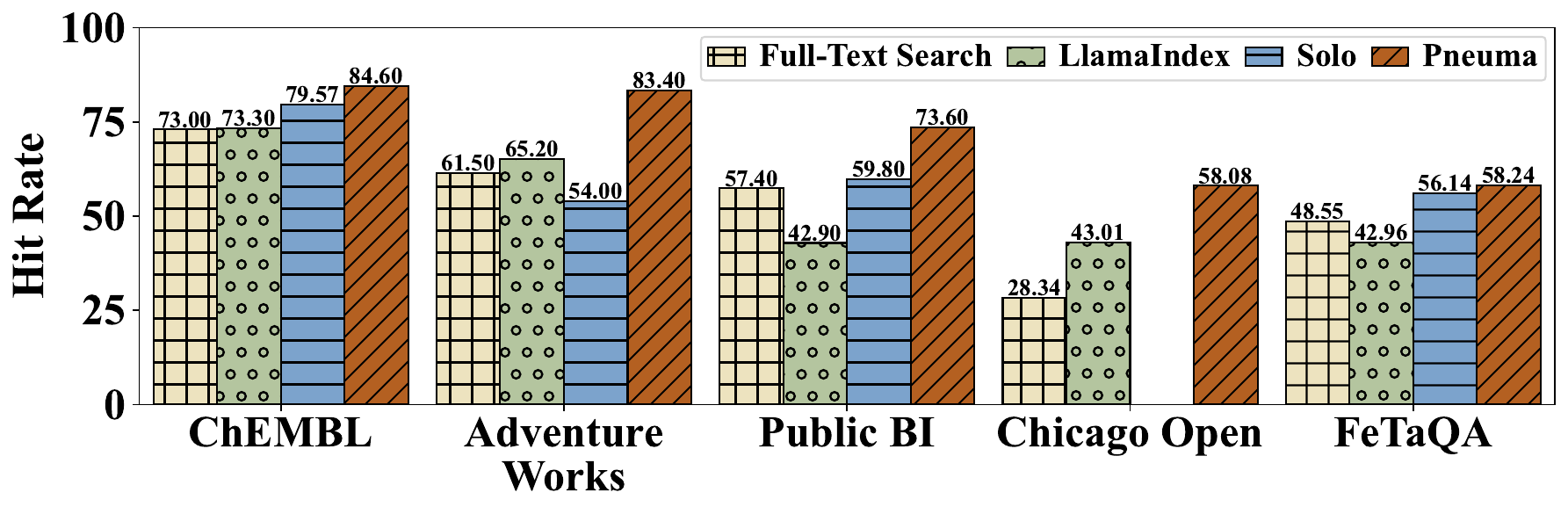} }}%
    \qquad
    \begin{minipage}[t]{0.39\linewidth}
        \centering
        \subfloat[\centering Hit Rates on BIRD (Original)]{{\includegraphics[width=\linewidth]{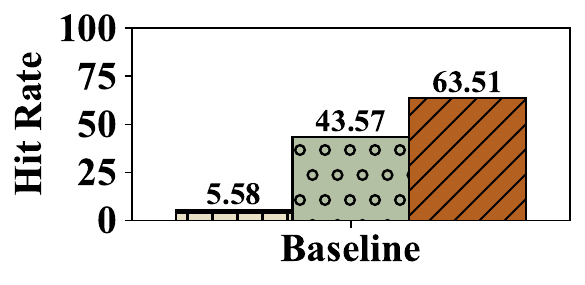} }}%
    \end{minipage}
    % \hfill
    \begin{minipage}[t]{0.39\linewidth}
        \centering
        \subfloat[\centering Hit Rates on BIRD (Annotated)]{{\includegraphics[width=\linewidth]{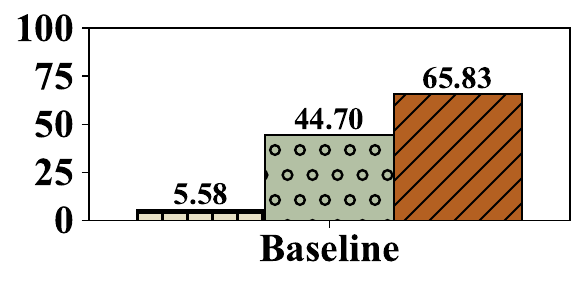} }}%
    \end{minipage}
    \caption{Hit Rates on Content Benchmarks and BIRD ($k=1$)}
    \label{fig:rq1_content}
\end{figure}

We compare \sys{} with the baselines on content benchmarks only (1k questions) across different datasets. As shown in Figure~\ref{fig:rq1_content}(a), \sys{} has higher hit rates than all baselines in every scenario. 
Other baselines have certain scenarios in which they perform worse compared to others. \fts performs worst on the Chicago dataset, \rag performs worst on the Public BI dataset, and \solo performs worst on the Adventure Works dataset. 

We also evaluate \sys{} on BIRD where 1595 questions meet our criteria of being fully answerable by a single table. We first do not tamper with the original BIRD benchmark, using the same 1595 questions and their tables to form BIRD (original). To account for questions that could potentially be answered by alternative tables with similar information, we created BIRD (annotated) by annotating all possible answer tables using \sys’s benchmark generator, identifying 120 (7.5\%) questions with two or more valid tables.

We exclude \solo{} because it generates an excessive number of vectors (8.88 billion vectors, $\sim$12TB storage). We limit \rag{} to only consider the first 5000 rows from each table in BIRD due to computational limitations: our instance only has access to approximately 90 GB of RAM, which is insufficient for \rag{}'s processing of the whole dataset because it needs to load all documents to the memory at once before converting them to embeddings. Nevertheless, \rag{} indexed 1.3 million vectors and took 34 hours to answer 1595 questions. In contrast, \sys{} did the same task in less than 13 minutes.

As shown in Figure~\ref{fig:rq1_content}(b), \sys{} outperforms other baselines significantly on BIRD (original). \sys{} achieves a hit rate of 63.51\%, which is 57.93 and 19.94 percentage points higher than \fts{} and \rag{}, respectively.  On BIRD (annotated), \sys{} achieves an even higher hit rate of 65.83\%, exceeding \fts{} and \rag{} by 60.25 and 21.13 percentage points, respectively. \fts{} performs badly mainly because lexical matching on row information is insufficient to answer the questions. \rag{} also utilizes row information, but the embedding model encodes semantic information, helping RAG better answer the questions.

\subsubsection{\sys{}'s Hit Rate in Answering Context Questions}

\begin{wrapfigure}{r}{0.55\textwidth}
    \centering
    \includegraphics[width=\linewidth]{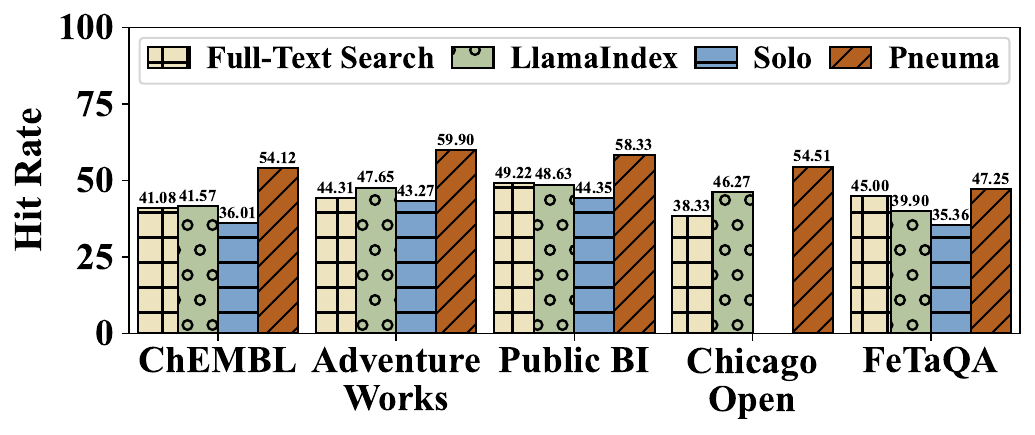}
    \caption{Hit Rates on Context Benchmarks ($k=1$)}
    \label{fig:rq1_context}
\end{wrapfigure}

In this section, we evaluate \sys{} and the baselines on context benchmarks only across different datasets (1k questions). As shown in Figure~\ref{fig:rq1_context}, \sys{} performs better than the three other baselines, highlighting \sys{}'s ability to handle a variety of context questions. For reference, \solo cannot index dataset context, but it is specifically trained on each dataset. Thus, it can point to the relevant tables for context questions such as the type of instances available in the tables and what each instance is about.

\subsubsection{Summary of Results}
Overall, we see that \sys{} is capable of handling a variety of context and content questions, outperforming all baselines across all datasets. On BIRD, the performance gap between \sys{} and other baselines is even more pronounced.

\subsection{RQ2: Efficiency of \sys{}}
\label{subsec:performance}

We compare the efficiency of \sys{} with \rag{} and \solo. \fts{} is excluded from this comparison because its performance suffers when there is insufficient lexical overlap between the question and the tables, and it lacks semantic understanding, which is often the performance bottleneck.
The efficiency is evaluated from two perspectives: \textbf{1. Online stage}: How quickly can \sys serve queries? This is measured in query throughput. \textbf{2. Offline stage}: How much time does \sys take to prepare data and how much is its storage footprint? We use the largest dataset, FeTaQa, to answer this research question. We evaluated performance using varying table counts: 625, 1250, 2500, 5000, and 10330 (the full FeTaQa dataset).

\begin{wrapfigure}{l}{0.55\textwidth}
    \centering
    \includegraphics[width=\linewidth]{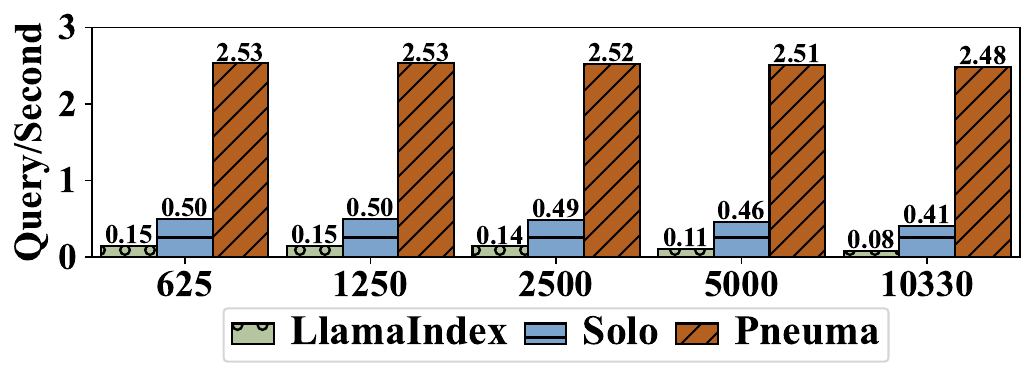}
    \caption{Query Throughput of \sys{}, \rag, and \solo ($k=1$)}
    \label{fig:rq2_queryThroughput}
\end{wrapfigure}

\subsubsection{Querying Throughput (Online)} We use \textbf{query throughput} to assess online efficiency. We compiled a list of 100 questions from our benchmarks and recorded the total time required to answer them. To ensure we did not capture noise in the results and that the results were consistent, we ran the performance benchmark for each setting 10 times and averaged the results. As shown in Figure~\ref{fig:rq2_queryThroughput}, \sys{} significantly outperforms both \rag{} and \solo{}. With 625 datasets, \sys{} is \textbf{16.6}$\times$ faster than \rag{} and \textbf{5}$\times$ faster than \solo{}. At 10,330 datasets, \sys{} outperforms \rag{} by \textbf{29.6}$\times$ and Solo by \textbf{6}$\times$. \sys{} is significantly more efficient and requires far less storage compared to \rag{}. This is because \sys{} generates concise table representations, indexing fewer documents and resulting in a smaller search space, whereas \rag{} must search through a larger number of vectors, increasing both computational and storage overhead.

\subsubsection{Data Preparation Time vs. Storage Footprint (Offline)}
\label{offline_time_storage}

\begin{figure}[b]
    \centering
    \includegraphics[width=0.9\linewidth]{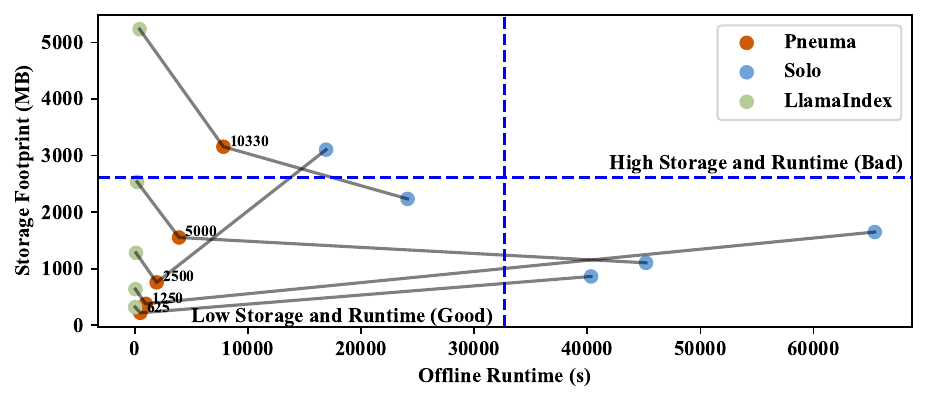}
    \caption{Comparison of Offline Runtime vs. Storage Footprint of \sys{}, \rag, and \solo}
    \label{fig:rq2_offlineVsStorage}
\end{figure}

The input dataset must be prepared before the system can answer questions. The preparation process for \sys and \rag{} includes ingesting the data, representing tables, and creating the index. For Solo, this process includes indexing the data and training a model. We call this offline time.  Figure~\ref{fig:rq2_offlineVsStorage} shows \rag{} has the fastest runtime but the largest storage footprint. \sys balances these factors, with linear growth in both.

The \rag{} system has very low offline runtime because the only data preparation needed is ingesting and directly generating embeddings. \sys{}, on the other hand, in addition to ingesting the data and generating embeddings, has an additional step of generating schema narrations. \sys takes considerably more preparation time because generating schema narrations requires prompting an LLM. However, this means \sys{} does not keep all row data from the dataset; thus, the storage footprint is smaller than \rag{}. With 625 tables, \sys requires \textbf{14.6}$\times$ longer but requires only \textbf{0.68}$\times$ the storage space. At 10330 tables, \sys requires \textbf{18.6}$\times$ longer but only \textbf{0.6}$\times$ the storage space. Compared to \solo, however, \sys is always faster.

The FeTaQA dataset predominantly consists of small tables, averaging 14 rows each. \sys{} creates a summary of 5 rows and an LLM-generated schema, while \rag indexes entire tables. Consequently, \sys's summaries only halve the storage size for FeTaQA.
However, on the Adventure Works dataset (88 tables, averaging 9,127 rows), \sys outperforms \rag significantly in both storage footprint and offline time. With this dataset, \rag{} requires 3357.2s offline time and 14.5GB of storage space, while \sys only requires 97.7s offline and 30.3MB. \rag{} takes \textbf{34.3}$\times$ longer and \textbf{477.7}$\times$ the storage footprint. 

For reference, \solo's data is irregular for 2 reasons. First, the training time, which consists of constructing training data and model training, varies. These steps depend on the number of questions in the training data instead of the number of tables in a dataset. Second, \solo's index size can be different depending on the number of vectors it generates. When it is < 1 million, an exact exhaustive search is used and vectors are stored without compression. Otherwise, an approximate search is used and vectors are compressed with product quantization. This causes \solo's storage footprints for 5,000 and 10,330 tables to be smaller than in other scenarios.

\subsubsection{Schema Summary Generation Throughput} In the event of changes in table schema or the addition of new tables, schema summaries need to be (re)generated. This process involves prompting the LLM to describe each table column that has changed. We report schema summary generation throughput in columns per second. A single process of \sys's summarizer can process \textbf{8 columns/second}. Whether processing 8 columns/second is sufficient depends entirely on the rate of schema changes in the dataset. If the rate of changes exceeds this throughput, it is easy to increase throughput by adding more processors. Further improvements to a processor's performance will also bring benefits.

\subsection{RQ3: Ablation Study}
\label{subsec:ablation_study}
We evaluate the impact of \sys{}'s components on its hit rates. Specifically, we investigate the contributions of \textsf{Table Summarizer}, \textsf{Hybrid Retrieval}, and \textsf{LLM Judge} by considering alternative variations for each of them.

\subsubsection{Impact of \textsf{Table Summarizer}}
\label{subsubsec:value_pneuma_summarizer}

\begin{wrapfigure}{r}{0.55\textwidth}
    \centering
    \includegraphics[width=\linewidth]{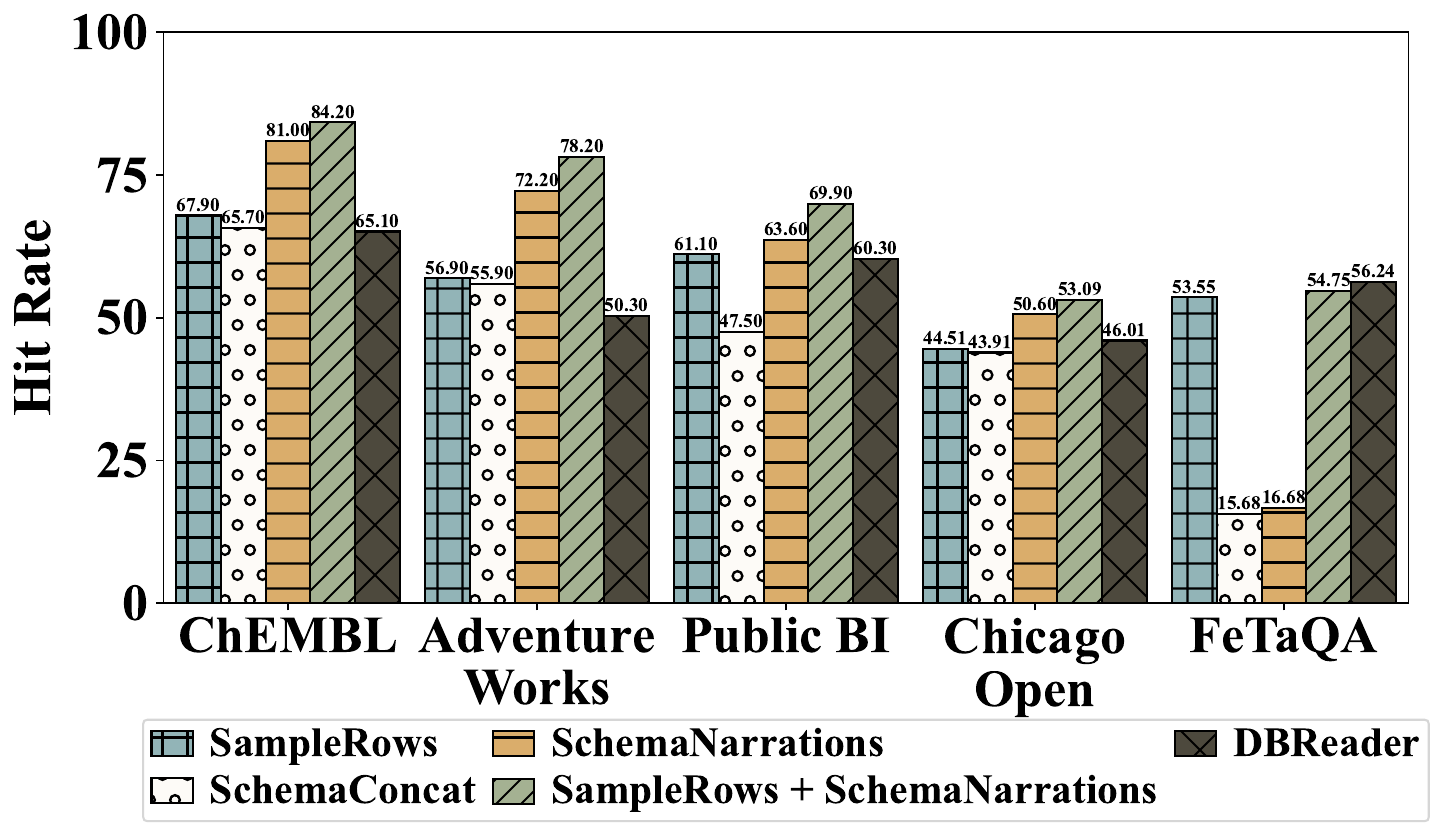}
    \caption{Hit Rates of \textsf{Hybrid Retrieval} on Content Benchmarks as its Content Summaries Vary}
    \label{fig:vary_content_summaries}
\end{wrapfigure}

To justify the design of \textsf{Table Summarizer}, we index variants of content summaries for \textsf{Hybrid Retrieval} ($k=1,n=5,\alpha=0.5$) and run it on the content benchmarks. The variants include \textbf{(1) \textsc{SampleRows}}: A random sample of $r=5$ rows from each table. For each row, we concatenate its values along with the corresponding column names. \textbf{(2) \textsc{SchemaConcat}}: A concatenation of the column names. \textbf{(3) \textsc{SchemaNarrations}}: Similar to \textsc{SchemaConcat}, but each column name is appended with LLM-narrated descriptions. \textbf{(4) \textsc{DBReader}}: Similar to \textsc{SampleRows}, but includes all rows.

\begin{wrapfigure}{l}{0.55\textwidth}
    \centering
    \includegraphics[width=\linewidth]{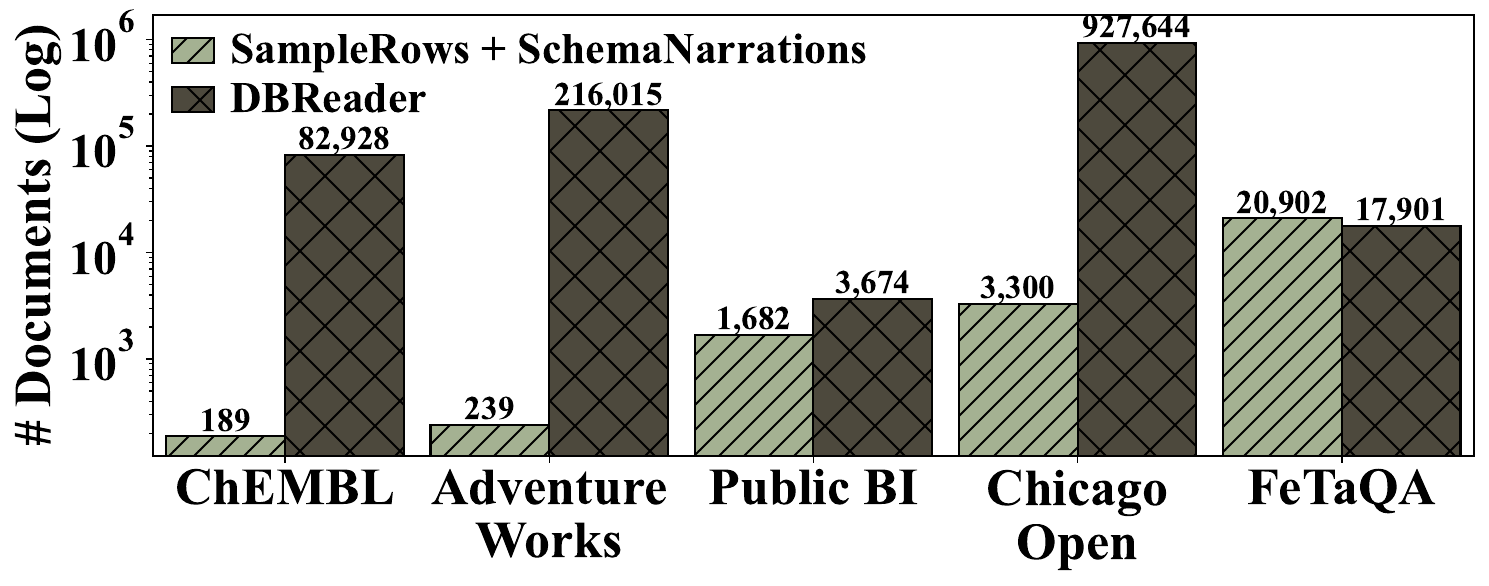}
    \caption{Comparison of Total Number of Documents (Log Scale) Between \textsf{Table Summarizer} and DBReader}
    \label{fig:summaries_count_comparison}
\end{wrapfigure}

\mypar{\textsc{SchemaNarrations} is better than \textsc{SchemaConcat}} As shown in Figure~\ref{fig:vary_content_summaries}, \textsc{SchemaNarrations} leads to better hit rates than \textsc{SchemaConcat} in every scenario. This demonstrates the effectiveness of narrating the schemas of the tables using an LLM. However, both of them obtain low hit rates on the FeTaQA dataset.

\mypar{\textsc{SampleRows} and \textsc{DBReader} improve hit rates on FeTaQA} Using row information leads to 3$\times$ better hit rate on the FeTaQA dataset compared to \textsc{SchemaNarrations} and \textsc{SchemaConcat}. This is because the FeTaQA dataset has many tables with similar schemas. For example, there are $420$ tables about elections with exactly the same schema. Thus, \textsf{Hybrid Retrieval} needs row-level information to differentiate tables with similar schemas.

Between \textsc{Samplerows} and \textsc{DBReader}, the differences in hit rates are minimal, but the latter corresponds to many more documents. For example, \textsc{DBReader} has 927,644 documents in the Chicago variant, which is 459$\times$ more documents than \textsc{Samplerows}. However, it only achieves $3.37\%$ better hit rate compared to \textsc{SampleRows}.

\mypar{\textsf{Table Summarizer} combines schema and row information} Although \textsc{Samplerows} and \textsc{DBReader} achieve much better hit rates on the FeTaQA dataset, \textsc{SchemaNarrations} is better on the other datasets. For instance, in the ChEMBL variant, \textsf{Hybrid Retrieval} obtains a hit rate of $81\%$, which is $13.1$ and $15.9$ percentage points higher than \textsc{SampleRows} and \textsc{DBReader}, respectively.

Therefore, \textsf{Table Summarizer} combines both schema narrations and sample rows as content summaries. In most scenarios, this approach leads to better hit rates and lower number of summaries---the latter is shown in Figure~\ref{fig:summaries_count_comparison}. In the Chicago dataset, for instance, \textsc{DBReader} has 281$\times$ more documents than \textsf{Table Summarizer}, but \textsf{Table Summarizer}'s hit rate is 5.29 percentage points higher.

\subsubsection{Impact of \textsf{Hybrid Retrieval}}
\label{subsubsec:value_pneuma_retriever}
In this section, we want to show that \textsf{Hybrid Retrieval} in \sys{}'s retriever ($k=1,n=5,\alpha=0.5$), is better than vector-search-only (\textsc{Pneuma-VS}) or full-text-search-only (\textsc{Pneuma-FTS}). All retrievers index the same documents: content summaries and table contexts. As shown in Table~\ref{tab:retrievers_comparison}, \textsf{Hybrid Retrieval} obtains the highest hit rates in $70\%$ of the scenarios and competitive otherwise. For example, in the Adventure Works-based content benchmark, \textsf{Hybrid Retrieval} obtains a hit rate of 81.40\%, which is 7 and 13.7 percentage points higher than \textsc{Pneuma-FTS} and \textsc{Pneuma-VS}, respectively.

\textsc{Pneuma-FTS} outperforms \textsc{Pneuma-VS} on content questions, while the opposite is true for context questions. Context questions are closely tied to their original contexts, so rephrasing has a greater impact on hit rates for these questions. \textsf{Hybrid Retrieval} leverages the fact that the two retrievers answer different sets of questions incorrectly, but with some overlap. It combines both retrievers and harnesses their respective strengths, resulting in a compounding improvement in performance.

\begin{table}[ht]
  \caption{Hit Rate Comparison Between \textsc{Pneuma-FTS}, \textsc{Pneuma-VS}, and \textsf{Hybrid Retrieval} ($k=1,n=5,\alpha=0.5$)}
  \centering
  \begin{tabular}{||c|c|c|c|c||}
    \hline
    \textbf{Dataset}  & \textbf{Benchmark} & \multicolumn{3}{c||}{\textbf{Retriever}} \\
    \cline{3-5}
                       & \textbf{Type}& \textsc{\textbf{Pneuma-FTS}} & \textsc{\textbf{Pneuma-VS}} & \textsf{\textbf{Hybrid Retrieval}} \\
    \hline\hline
    ChEMBL   & Content & 75.40 & 73.00 & \textbf{83.00}\\
    \cline{2-5}
             & Context & 44.22 & \textbf{53.24} & 50.88\\
    \hline
    Adventure   & Content & 74.40 & 67.70 & \textbf{81.40} \\ 
    \cline{2-5}
    Works       & Context & 48.24 & \textbf{57.55} & 56.08 \\     
    \hline
    Public BI    & Content & \textbf{72.10} & 42.90 & 71.10 \\
    \cline{2-5}
                 & Context & 49.80 & 57.65 & \textbf{57.75} \\
    \hline
    Chicago      & Content & 50.10 & 40.92 & \textbf{55.09}\\
    \cline{2-5}
    Open         & Context & 41.18 & 57.16 & \textbf{52.25} \\                
    \hline
    FeTaQA       & Content & 54.05 & 25.17 & \textbf{56.14}\\
    \cline{2-5}
                 & Context & 40.10 & 44.90 & \textbf{46.96}\\                 
    \hline
  \end{tabular}
  \label{tab:retrievers_comparison}
\end{table}

\subsubsection{Impact of \textsf{LLM Judge}}
\label{subsubsec:value_pneuma_reranker}
To gain insight into \textsf{LLM Judge}'s impact on \textsf{Hybrid Retrieval}, we compare it with two pre-trained re-ranker models: BAAI/bge-reranker-v2-m3 (BGE)~\cite{BGE2024} and dunzhang/stella\_en\_1.5B\_v5 (Stella)~\cite{kusupati2022stella}.  The former is a widely-used re-ranker model, while the latter is the best embedding model for re-ranking purposes for its size (as of 2024-10-17) according to the MTEB Leaderboard~\cite{MTEB2023}. (It is overall ranked second, slightly below a model 7 times its size.) At the same time, we try a different model for \textsf{LLM Judge}: HuggingFaceH4/zephyr-7b-beta (Zephyr)~\cite{Zephyr2023}.

\begin{figure}[h!]
    \centering
    \subfloat[\centering Hit rates on content benchmarks]{{\includegraphics[width=0.48\linewidth]{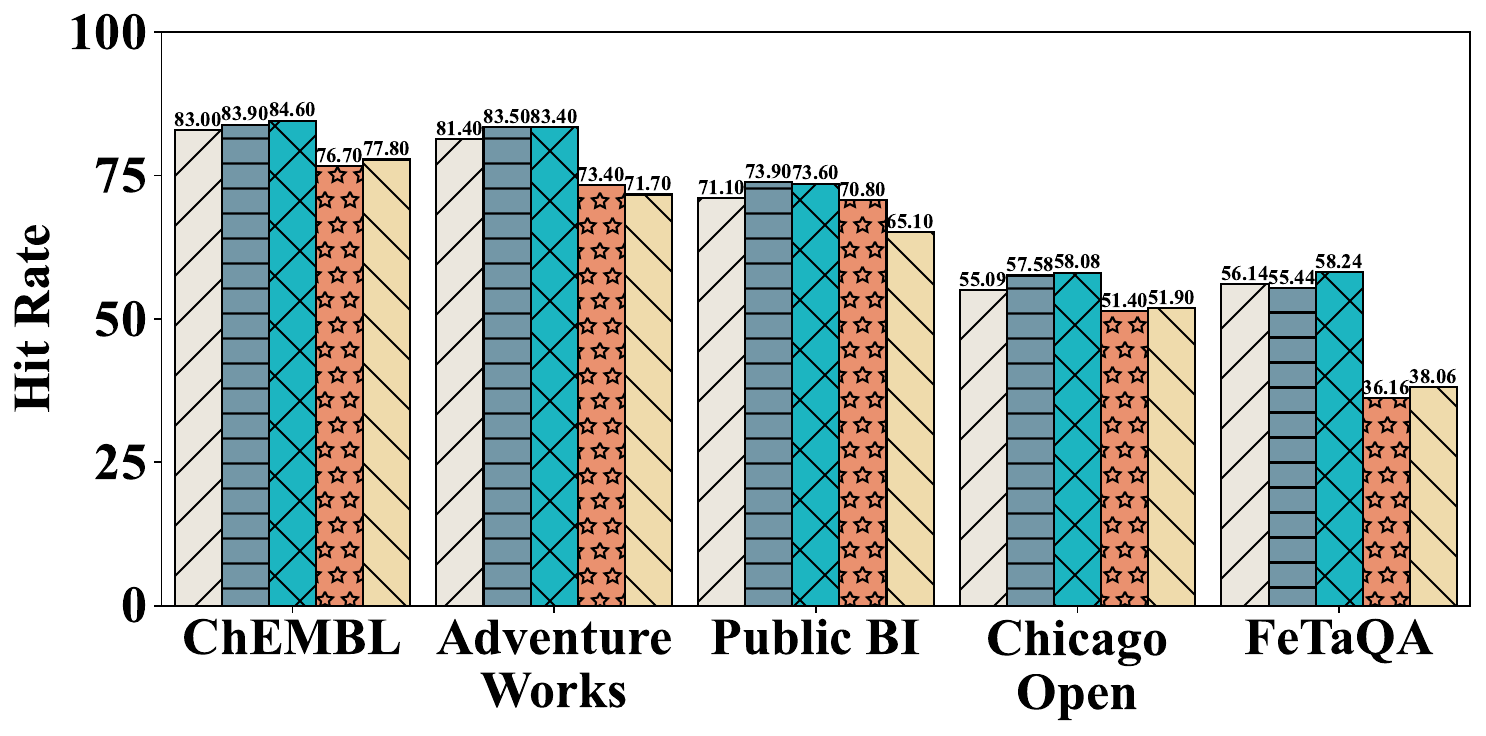} }}%
    % \qquad
    \subfloat[\centering Hit rates on context benchmarks]{{\includegraphics[width=0.48\linewidth]{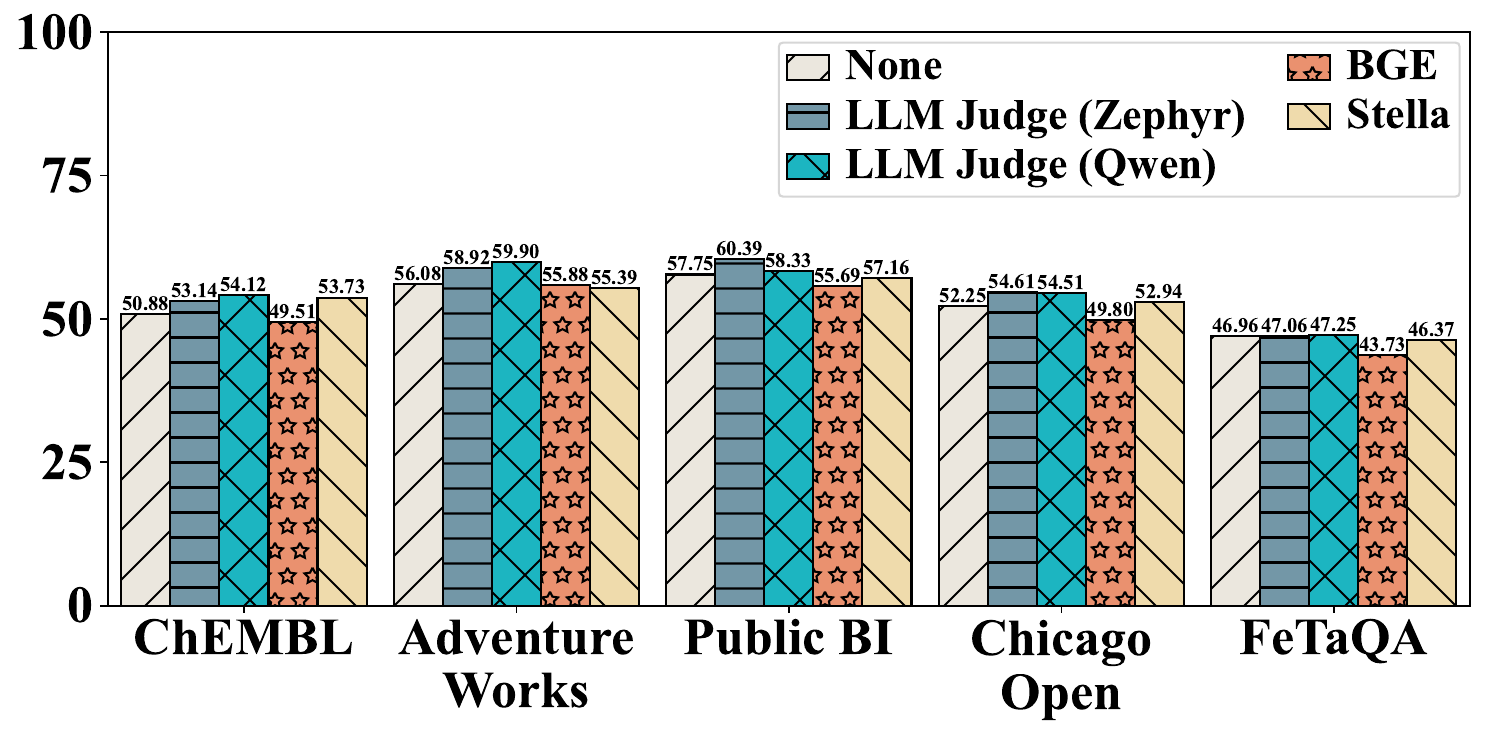} }}%
    \caption{Hit Rates of \textsf{Hybrid Retrieval} ($k=1,n=5,\alpha=0.5$) with Different Judges}
    \label{fig:reranker_comparison}
\end{figure}

We compare the hit rates of \textsf{Hybrid Retrieval} ($k=1,n=5,\alpha=0.5$), with and without judges, in Figure~\ref{fig:reranker_comparison}. We observe that \textsf{LLM Judge}, both Zephyr and Qwen variants, perform better than the re-ranker models. For example, in the FeTaQA-based content benchmark, \textsf{LLM Judge} (Qwen) obtains a hit rate of 58.24\%, which is 22.08 and 20.18 percentage points higher than BGE and Stella, respectively. In almost all scenarios, both Stella and BGE reduce hit rates. This is undesirable because users have traded query time for better hit rates, not worse. Between Zephyr and Qwen, the differences are subtle, with Qwen obtaining slightly better hit rates in more scenarios. Overall, \textsf{LLM Judge}'s performance is not sensitive to the selection of LLMs of similar size.

\subsection{RQ4: Microbenchmarks}
\label{subsec:microbenchmarks}
We evaluate \sys's retrieval performance by varying its hyperparameters ($\alpha$, $k$, and $n$) and testing different LLMs as schema summarizers for robustness. We also explore the baselines' behavior when questions share many keywords with dataset contexts and row values, and investigate the impact of hallucinations.

\subsubsection{Changing $\alpha$}
\label{subsubsec:change_alpha}
\begin{figure}[h]
    \centering
    \subfloat[\centering Hit rates on content benchmarks with a variety of $\alpha$ values]{{\includegraphics[width=0.9\linewidth]{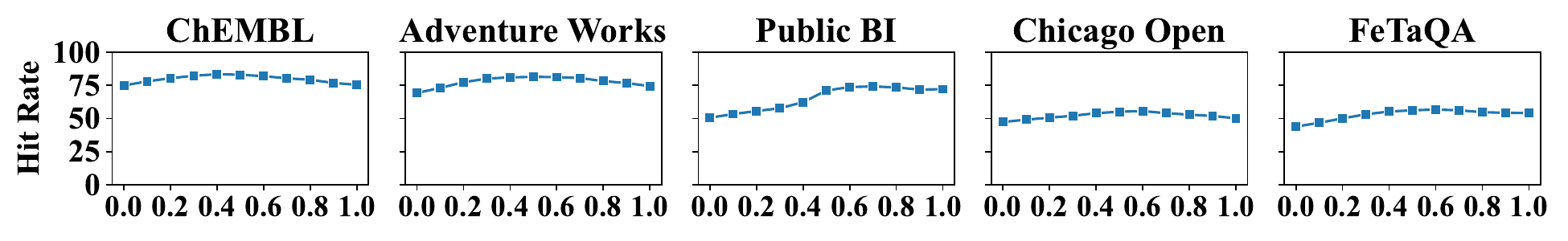} }}%
    \qquad
    \subfloat[\centering Hit rates on context benchmarks with a variety of $\alpha$ values]{{\includegraphics[width=0.9\linewidth]{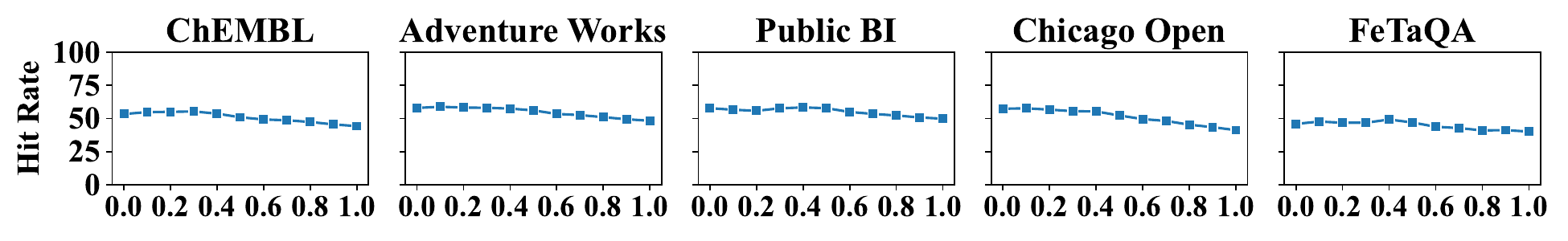} }}%
    \caption{Hit Rates of \textsf{Hybrid Retrieval} ($k=1,n=5$) on Content and Context Benchmarks as $\alpha$ Varies}
    \label{fig:vary_alpha}
\end{figure}

This hyperparameter linearly combines the normalized relevance scores given by vector and full-text retrievers. Higher $\alpha$ gives more weight to the latter. To understand the impact, we experiment with $\alpha \in \{0.0,0.1,...,1.0\},k=1,n=5$. As shown in Figure~\ref{fig:vary_alpha}, the hit rates on both benchmarks peak at around $0.4-0.6$. This suggests the effectiveness of weighting both retrievers at approximately the same weight.

\subsubsection{Changing k}
\label{subsubsec:effect_k}
\begin{figure}[h]
    \centering
    \subfloat[\centering Hit rates on content benchmarks with a variety of k values]{{\includegraphics[width=0.9\linewidth]{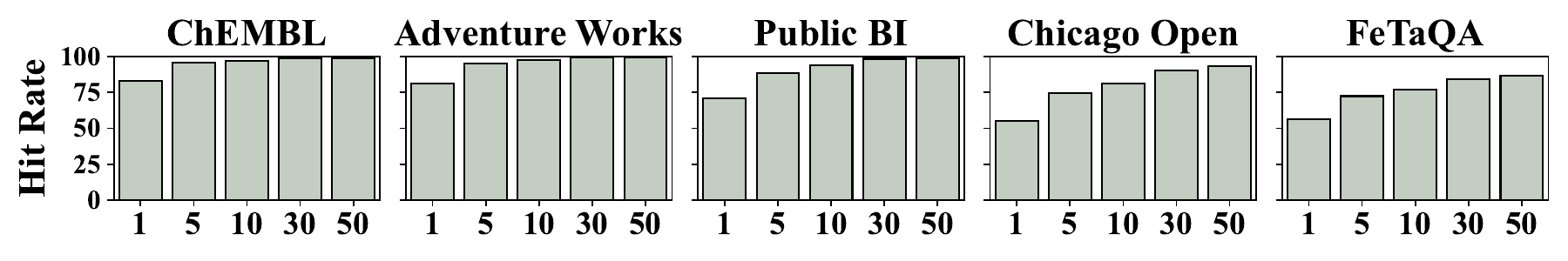} }}%
    \qquad
    \subfloat[\centering Hit rates on context benchmarks with a variety of k values]{{\includegraphics[width=0.9\linewidth]{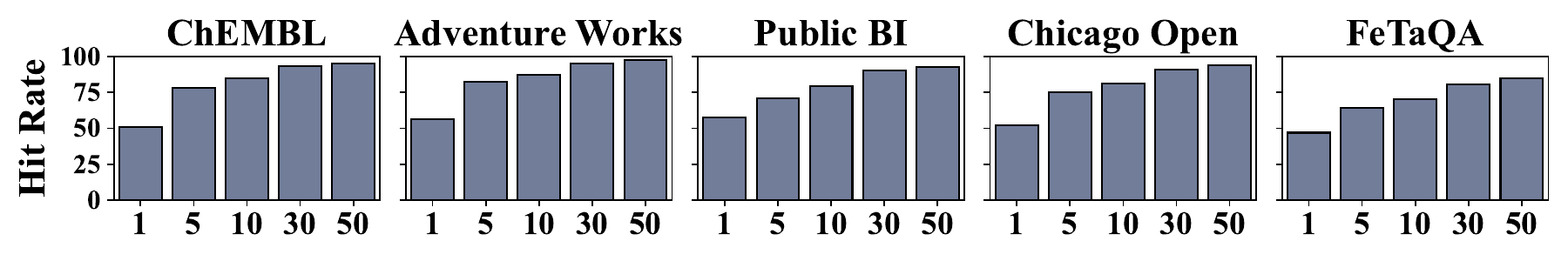} }}%
    \caption{Hit Rates of \textsf{Hybrid Retrieval} ($n=5,\alpha=0.5$) on Content and Context Benchmarks as k Varies}
    \label{fig:vary_k}
\end{figure}

$k$ determines the number of documents retrieved. Higher $k$ values increase the chance of retrieving relevant tables, potentially improving hit rates. We tested $k \in \{1,5,10,30,50\}$ with $n=5,\alpha=0.5$. Figure~\ref{fig:vary_k} confirms this: hit rates improve for both benchmarks as $k$ increases, with the largest gain from $k=1$ to $k=5$, followed by smaller but continued improvements.

\subsubsection{Changing n}
\label{subsubsec:effect_n}
\begin{figure}[h]
    \centering
    \subfloat[\centering Hit rates on content benchmarks with a variety of n values]{{\includegraphics[width=0.9\linewidth]{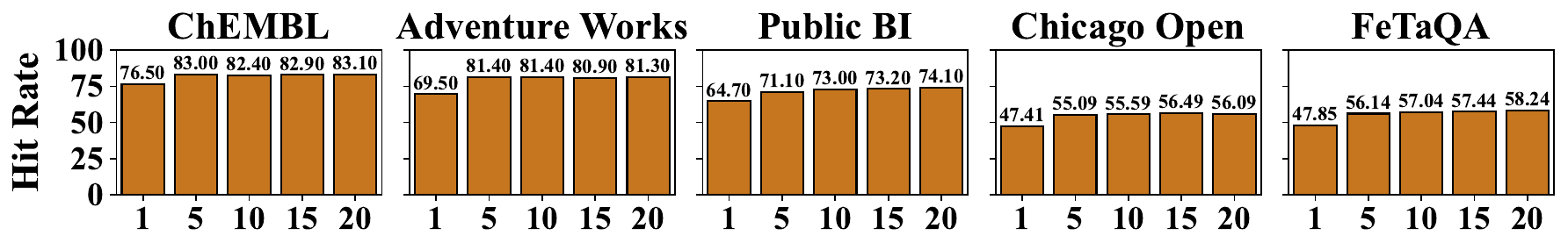} }}%
    \qquad
    \subfloat[\centering Hit rates on context benchmarks with a variety of n values]{{\includegraphics[width=0.9\linewidth]{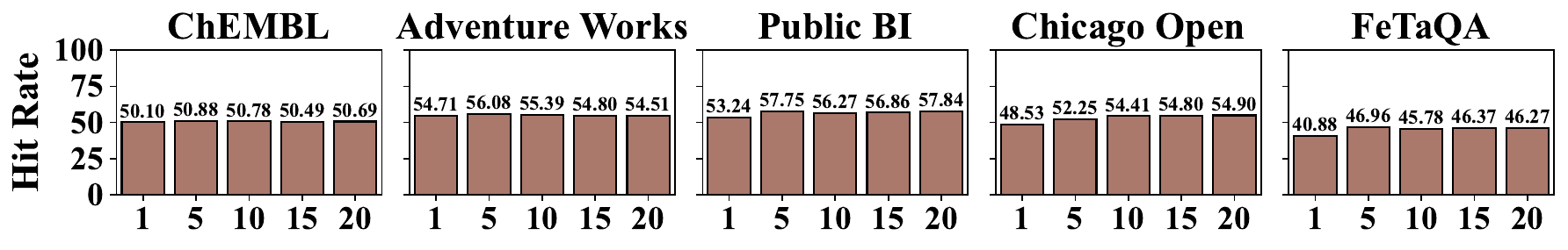} }}%
    \caption{Hit Rates of \textsf{Hybrid Retrieval} ($k=1,\alpha=0.5$) on Content and Context Benchmarks as n Varies}
    \label{fig:vary_n}
\end{figure}

The last hyperparameter is the multiplicative factor of $k$ for retrieving documents from both vector and full-text indices. By retrieving $nk$ instead of $k$ documents, we gather more relevant ones. We study the effect of changing the value of $n$ on hit rates. We do so by experimenting with these hyperparameters: $k=1,n \in \{1,5,10,15,20\},\alpha=0.5$. As shown in Figure~\ref{fig:vary_n}, the differences in hit rates occur mostly from $n=1$ to $n=5$, and are minuscule otherwise.

\subsubsection{Changing LLMs for Schema Summarizer}
\label{subsubsec:effect_llm}

\begin{wrapfigure}{r}{0.55\textwidth}
    \centering
    \includegraphics[width=\linewidth]{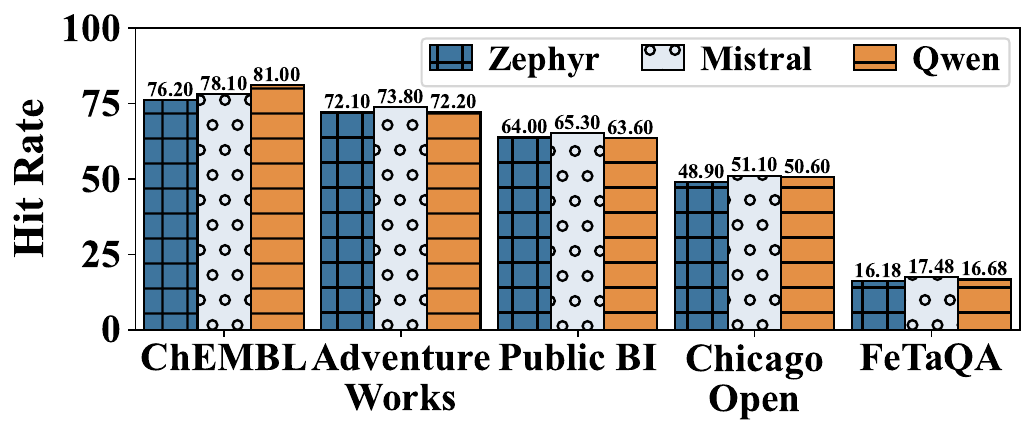}
    \caption{\textsf{Hybrid Retrieval} Hit Rates ($k=1,n=5,\alpha=0.5$) on Content Benchmarks as Schema Summaries Vary}
    \label{fig:vary_llm}
\end{wrapfigure}

We tested Mistral-7B-Instruct-v0.3 and Zephyr-7b-beta as alternatives to Qwen for schema summarization, using \textsf{Hybrid Retrieval} ($k=1,n=5,\alpha=0.5$). Figure~\ref{fig:vary_llm} shows minimal differences in hit rates, with Mistral slightly outperforming others. We retain Qwen for its conciseness, enabling faster summarization. These results demonstrate that our system's performance is not significantly model-dependent.

\subsubsection{Investigating the impact of hallucinations}
\label{subsubsec:effect_llm_config}

We have made efforts to minimize hallucinations by prompting the LLM to avoid generating descriptions when uncertain and using greedy decoding. Despite our efforts, certain LLMs still hallucinate. While we expect this to improve as LLM technology advances, we conducted experiments to assess how hallucination-prone summaries affect hit rates. Specifically, we compared the hit rates of \textsf{Hybrid Retrieval} ($k=1, n=5, \alpha=0.5$) when indexing schema summaries generated with three different LLM configurations: \textbf{(1) \textsc{Temp-0}} (default: greedy search), \textbf{(2) \textsc{Temp-1.5}} (sampling enabled with temperature set to 1.5), and \textbf{(3) \textsc{Non-Instruct-Temp-1.5}} (same as (2) but using the non-instruct variant of Qwen).

\begin{wrapfigure}{l}{0.55\textwidth}
    \centering
    \includegraphics[width=\linewidth]{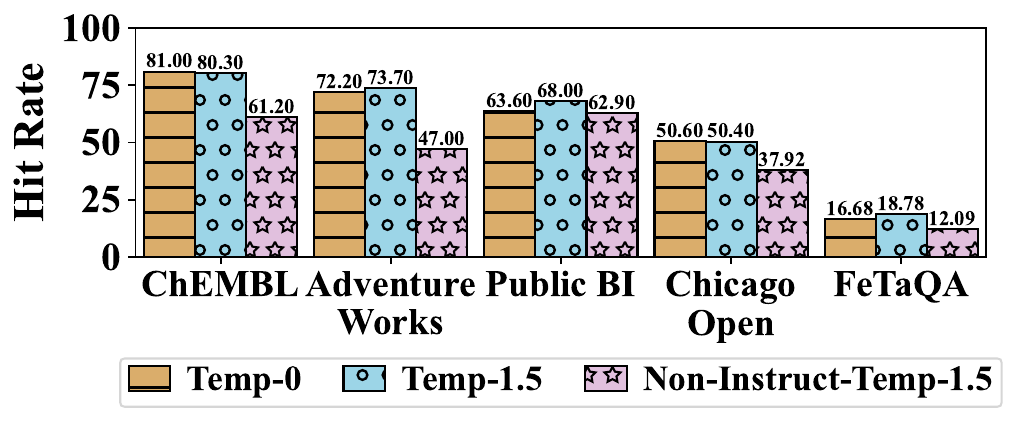}
    \caption{Hit Rates of \textsf{Hybrid Retrieval} ($k=1,n=5,\alpha=0.5$) on Content Benchmarks for different LLMs}
    \label{fig:vary_llm_2}
\end{wrapfigure}

Figure~\ref{fig:vary_llm_2} shows that \textsc{Temp-0} performs on par with \textsc{Temp-1.5} because increasing the temperature results in summaries with equivalent meanings but a broader vocabulary. In contrast, \textsc{Non-Instruct-Temp-1.5} produces hallucinated summaries. For example, the 'molregno' column in the ChEMBL dataset represents an internal ID for compounds. Both \textsc{Temp-0} and \textsc{Temp-1.5} correctly identify it as IDs for molecule or compound, but \textsc{Non-Instruct-Temp-1.5} incorrectly describes it as representing alphanumeric molecules. Similar inaccuracies are observed across other summaries generated by \textsc{Non-Instruct-Temp-1.5}, leading to significantly lower hit rates.

To assess whether the LLM summarizer is prone to hallucinations, \sys's benchmark generator can be used to create a table discovery benchmark for a sample of tables. The benchmark serves as a test dataset to evaluate the quality of the generated summaries. A very low hit rate on this test set indicates that the LLM is prone to hallucinations and fails to produce useful summaries.

\subsubsection{Using Keyword-Heavy Benchmarks}

\begin{figure}[h]
    \centering
    \subfloat[\centering Hit Rates on Keyword-Heavy Content Benchmarks]{{\includegraphics[width=0.48\linewidth]{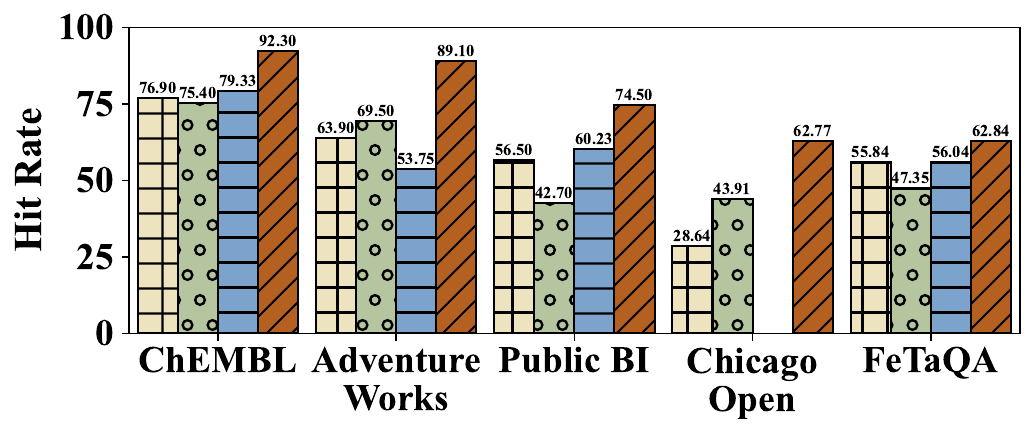} }}
    \subfloat[\centering Hit Rates on Keyword-Heavy Context Benchmarks]{{\includegraphics[width=0.48\linewidth]{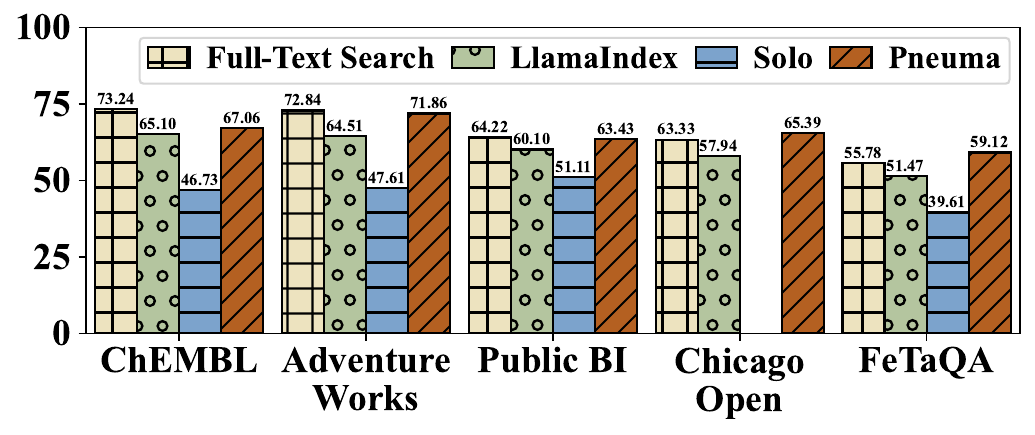} }}%
    \caption{Hit Rates on Keyword-Heavy Content and Context Benchmarks ($k=1$)}
    \label{fig:keyword_heavy_benchmarks}
\end{figure}

We evaluated \sys{} ($k=1,n=5,\alpha=0.5$) and other baselines on keyword-heavy variants of context and content benchmarks, where questions share many keywords with contexts and row values.
Figure~\ref{fig:keyword_heavy_benchmarks} shows \sys{} outperforming all baselines on content benchmarks and competing closely with \fts{} on context benchmarks. Due to keyword overlap, \fts{} generally outperforms \rag, contrary to more realistic scenarios (Section~\ref{subsec:quality}). These results highlight \sys{}'s effectiveness in handling keyword-heavy questions, leveraging its consideration of lexical information.

\section{Related Work}
\label{sec:relatedwork}

\mypar{Information Retrieval} Information retrieval methods such as  BM25~\cite{robertson2009probabilistic} match queries with documents based on term overlap, while vector search captures semantic similarity by embedding queries and documents in high-dimensional spaces~\cite{karpukhin2020dense}. Hybrid approaches combine both methods to balance precision and semantic understanding~\cite{bruch2023analysis}. \sys{} extends these techniques with an LLM-judge which aids in reranking results. Unlike other LLM reranking methods~\cite{gao2024llm}, which rely on LLMs to directly output a ranking, \sys{} assigns the LLM a simpler task: binary classification of the relevance of documents to questions.

\mypar{Tabular Data Discovery} Data discovery investigates identifying and retrieving relevant data—either an individual dataset~\cite{wang2023solo, aurum, castelo2021auctus} or a combination of datasets~\cite{gong23, galhotra2023metam, fan2024}—to satisfy the user's information need. \sys{} focuses on retrieving relevant datasets individually, without performing any combination or integration across them. Keyword search~\cite{castelo2021auctus, zhang2018ad} is a traditional interface to discover tabular data. However, keywords are not specific enough to express user intent, and it is difficult for users to choose the right keywords.
Aurum~\cite{fernandez2018aurum} proposed programming APIs for more advanced data discovery features but requires users to have technical backgrounds and hence less accessible. Recently natural-language question (NLQ) \cite{herzig2021open, wang2021retrieving, wang2023solo} has become a more popular interface for tabular data discovery, thanks to the quick development of LLMs. Systems such as OpenDTR~\cite{herzig2021open} and GTR~\cite{wang2021retrieving} require expensive human-annotated data for training on a new table corpus to perform well. Although Solo~\cite{wang2023solo} proposed a self-supervised approach, it suffers from long training time as shown in Section~\ref{offline_time_storage}. \sys uses NLQ as an interface and does not need training by exploiting the combination of pre-trained embedding model, LLM, and hybrid search.
% , offering a cheaper data discovery solution using NLQ.

\mypar{Table Representation for Data Discovery}
There are generally two approaches to table representation for data discovery. The first approach learns representations on tabular data~\cite{herzig2021open, yin2020tabert, liu2021tapex, wang2021retrieving, zhao2022leva}. However, this approach may not generalize well to different datasets~\cite{wang2023solo}. The second approach converts tables into multiple text snippets and uses off-the-shelf models~\cite{huggingface_models} to represent them. Among text-based approaches, \citet{Liu_LlamaIndex_2022} converts each row to a snippet and then encodes each row snippet into a vector. \citet{wang2023solo} treats each pair of columns in a row as a text snippet. Both \citet{Liu_LlamaIndex_2022} and \citet{wang2023solo} generate many vectors, impacting storage and indexing scalability. \sys{} produces fewer vectors without trading search quality, and even performs better by representing tables with LLM-generated schema descriptions and sample rows.

\mypar{Benchmarks for Tabular Data Discovery} QATCH~\cite{QatchBenchmark}, BIRD~\cite{BirdBenchmark}, and Spider~\cite{Spider2024}, are designed for table question answering and text-to-SQL tasks. QATCH reveals the correct table in the question, making it unsuitable for table discovery. BIRD and Spider can be adapted for table discovery tasks but involve fewer tables, offering a less challenging setting compared to \textsc{Pneuma}'s benchmark. CMDBench~\cite{CMDBench} aligns better with \textsc{Pneuma}'s goals but lacks comprehensive ground truth, which is critical for accurately interpreting hit rates. \textsc{Pneuma} offers a more comprehensive and challenging benchmark generator for table discovery tasks.

\mypar{Table Question Answering}
Table question answering (QA) is a prominent downstream application of tabular data. Prior work has explored both specialized and general-purpose approaches. TaPaS~\cite{herzig-etal-2020-tapas} and OmniTab~\cite{jiang-etal-2022-omnitab}, for instance, train/fine-tune specialized models for table QA, while others use large language models with tools (e.g., SQL engines), as in Chain-of-Table~\cite{wang2024chainoftable} and ReAcTable~\cite{Zhang2024reactable}. LOTUS~\cite{patel2024semanticoperators} introduces semantic operators for transforming tables using natural-language criteria, which has been demonstrated to be useful for table QA~\cite{biswal2024tag}. Palimpzest~\cite{palimpzestCIDR} tackles analytical queries over input tables or tabulated unstructured data. This generalization is also adopted by Aryn~\cite{anderson2024aryn}, which handles tables, among other types of modality, in documents. Overall, these systems could serve as downstream users of the tables retrieved by \sys{}.

\section{Conclusions}
\label{sec:conclusions}

We introduce \sys{}, an end-to-end system that helps users discover desired tables in large data repositories using natural language queries based on both data content and context. \sys{} enables efficient table discovery through two key components: table representation and table retrieval. Table representation generates schema and row summaries by leveraging an LLM to narrate schemas, while table retrieval combines the \textsf{LLM Judge} with two traditional information retrieval methods, effectively utilizing the strengths of both approaches. \sys{} outperforms state-of-the-art table retrieval systems in both response quality and efficiency, while requiring orders of magnitude less storage.

\section{Acknowledgements}
\label{sec:acknowledgements}

We thank the reviewers for their constructive feedback. We thank the Indonesia-US Research Collaboration Open Digital Technology program, the Data Science Institute's Research Initiative on Data Ecology, and the National Science Foundation (CAREER Award 2340034) for supporting this project.

\clearpage

\bibliographystyle{ACM-Reference-Format}
\bibliography{main}

\end{document}